\documentclass[journal,10pt]{IEEEtran}

\usepackage{amsmath,amssymb,amsfonts}
\usepackage{algorithmic}
\usepackage{graphicx}
\usepackage{textcomp}
\usepackage{float}
\usepackage{fancyvrb} 
\usepackage{textcomp} 
\usepackage{enumitem} 
\usepackage{pifont} 
\def\BibTeX{{\rm B\kern-.05em{\sc i\kern-.025em b}\kern-.08em
    T\kern-.1667em\lower.7ex\hbox{E}\kern-.125emX}}

\usepackage{microtype}
\mathchardef\mhyphen="2D

\usepackage{atbegshi}
\AtBeginDocument{\addtocounter{page}{-1}\AtBeginShipoutNext{\AtBeginShipoutDiscard}} 

\begin{document}

\title{Formal Certification Methods for Automated Vehicle Safety Assessment}

















\author{Tong Zhao,~\IEEEmembership{Member, IEEE},
Ekim Yurtsever,~\IEEEmembership{Member, IEEE}, Joel A. Paulson,~\IEEEmembership{Member, IEEE}, and Giorgio Rizzoni,~\IEEEmembership{Fellow, IEEE}.
}


\thanks{T. Zhao is with the Department of Mechanical and Aerospace Engineering, The Ohio State University, Columbus, OH, 43212, USA e-mail: zhao.1991@osu.edu}

\thanks{E. Yurtsever is with the Department of Electrical and Computer Engineering, The Ohio State University, Columbus, OH, 43212, USA e-mail: yurtsever.2@osu.edu}

\thanks{J. Paulson is with the Department of Chemical and Biomolecular Engineering, The Ohio State University, Columbus, OH, 43212, USA e-mail: paulson.82@osu.edu}

\thanks{G. Rizzoni is with the Department of Mechanical and Aerospace Engineering, The Ohio State University, Columbus, OH, 43212, USA e-mail: rizzoni.1@osu.edu}


\thanks{This work has been submitted to the IEEE for possible publication. Copyright may be transferred without notice, after which this version may no longer be accessible.}


\maketitle

\begin{abstract}
Challenges related to automated driving are no longer focused on just the construction of such automated vehicles (AVs), but in assuring the safety of their operation. 
Recent advances in Level 3 and Level 4 autonomous driving have motivated more extensive study in safety guarantees of complicated AV maneuvers, which aligns with the goal of ISO 21448 (Safety of the Intended Functions, or SOTIF), i.e. minimizing unsafe scenarios both known and unknown, as well as Vision Zero -- eliminating highway fatalities by 2050. A majority of approaches used in providing safety guarantees for AV motion control originate from formal methods, especially reachability analysis (RA), which relies on mathematical models for the dynamic evolution of the system to provide guarantees. However, to the best of the authors' knowledge, there have been no review papers dedicated to describing and interpreting state-of-the-art of formal methods in the context of AVs. In this work, we provide both an overview of the safety verification, validation and certification process, as well as review formal safety techniques that are best suited to AV applications. We also propose a unified scenario coverage framework that can provide either a formal or sample-based estimate of safety verification for full AVs. 
Finally, remaining challenges and future opportunities beyond the scope of current published research for assured AV safety are presented. 
\end{abstract}

\begin{IEEEkeywords}
Automated vehicle, Verification and validation, Reachability analysis, Safety guarantees, Motion planning, Safety of the intended function, Scenario coverage
\end{IEEEkeywords}

\IEEEpeerreviewmaketitle

\section{Introduction}

Automated vehicle (AV) technology is expected to be massively deployed in the next decade or two, with multiple SAE J3016 level 2 automated driving systems developed and deployed over the last two decades\cite{vahidi2003research,2015_adas_survey}, and some level 3 system regionally approved and announced\cite{immen_2021,honda_2020}. As a daily means of transportation for humans and goods, more rigorous safety guarantee has to be investigated and eventually implemented in the automated fleet that will dominate the road traffic around 2050\cite{litman2017autonomous}. Recent poll suggests that safety is the primary concern for AV acceptance, compared to economic impact or privacy concern\cite{ho2021complementary}. And it is believed that by mass adoption of automated vehicles, road safety will be significantly improved compared to the road traffic status quo\cite{airbib2017rethinking}. The gap between this incoming autonomous future and current state-of-art safety measures is to be filled by continued effort of scientists and engineers, as well as legislators.

Previous surveys on AV technologies have been focusing on the various aspects of the technical challenges and implementation, such as ways to perform perception, decision making, vehicle control, human machine interaction\cite{yurtsever2020survey}, with a few reviews covering the issue of AV safety verification and validation\cite{koopman2017autonomous,kone2019safety,wang2020safety,tahir2020coverage,batsch2020taxonomy,riedmaier2020survey,kress2021formalizing,weng2021,sun_2021}.

With the rise of real word deployment of ADS systems, significant amount of verification and validation (V\&V) is required against the scale of scenarios in real world that ADS will encounter. A common strategy of V\&V to maximize scenario coverage is to verify ADS in virtual simulations and simulated large amount of generated scenario samples (see \cite{riedmaier2020survey,tahir2020coverage,sun_2021,ding2022survey} for detailed reference).

The challenge in scenario sampling-based V\&V is quantifying the amount of sample required to guarantee reasonable coverage, so that the risk created by ADS improper actuation can be curbed. A reasonable scenario coverage guarantee is also the prerequisite for type approval in recent legislation for highly automated vehicle (HAV) deployment\cite{germany_level3_2017,UN157}. An alternative set of methods based on formal verification however, address the scenario coverage from a different approach through specification satisfaction\cite{kress2021formalizing}. Because of the potential of formal methods in scenario coverage, emerging research have started to combing formal properties with control synthesis\cite{chen2018signal,leung2020infusing} and safety verification\cite{weng2021}.

In light of the need for ADS quantifiable verification, we propose a unified scenario coverage framework to address the unresolved pain point in AV scenario coverage. We provide a quantitative definition of unified safe scenario coverage given an accepted coverage-representation volume of a single scenario (for sample-based methods), or given so called ``specification penetration rate" of formal safety specifications (for formal methods). This quantitative definition unites sample-based methods and formal methods in the same quest to provide scenario coverage for safety verification. A candidate AV control policy can go through either formal or sample-based safety verification to be tested of its safe scenario coverage in a specified ODD. We compare the pros and cons of sample-based methods and formal methods based on how they reach this safety coverage and the associated costs and issues with each method.

The contributions of this paper is three-fold: (1) we provide a holistic overview of different aspects in ADS safety guarantee, verification, validation and certification; (2) we provide a unified safe scenario coverage framework for both sample-based and formal methods for safety verification; (3) we conduct a state-of-the-art literature review on formal methods for automated vehicle safety, and we categorize how in different ways formal methods (especially reachability analysis) can be used to provide safety verification or guarantee for motion control. Lastly, we point out trends and opportunities in the field of formal safety verification for ADS.


\begin{table*}[t]
    \centering
    \caption{Legislation status of automated vehicle operation}
    \begin{tabular}{p{0.05\textwidth}p{0.05\textwidth}p{0.15\textwidth}p{0.15\textwidth}p{0.12\textwidth}p{0.12\textwidth}p{0.2\textwidth}}
        \hline\hline\\
        Region & Testing & \multicolumn{2}{c}{Deployment Legislation} & \multicolumn{2}{c}{Safety Driver Removal} & Approved Deployment Case(s)\\
        {} & {} & Level 3 & Level 4 & Testing & Deployment & {}\\
        \hline\\
        US & \checkmark* & \multicolumn{2}{p{0.25\textwidth}}{State-specific** (SAE J3016 classification\cite{J3016_202104} in 2014)} & Conditional*** & Conditional**** & Nuro (California, Dec 2020)\cite{nuro_2020}; Waymo, Cruise (California, Sep 2021)\cite{waymocruise_2021}\\
        EU & \checkmark & UNECE No.79, Apr 2021\cite{un_level3steer_2020}; Germany, Jun 2017\cite{germany_level3_2017} & Germany, Jul 2021\cite{schmid_2021} & Germany, 2022\cite{germany_safetydriver_2022} & - & Mercedes-Benz (Level 3 ALKS, Dec 2021)\cite{mercedes_2021}\\
        UK & \checkmark & Spring 2022\cite{uk_level3_kickoff,uk_level3_2022} & - & - & - & -\\
        China & \checkmark & \multicolumn{2}{p{0.25\textwidth}}{General legislation, Apr 2021\cite{mallesons_2021}, AV level classification, Mar 2022\cite{china_taxonomy_2022}} & Conditional & - & Robotaxi w/ safety driver: Baidu, Pony.ai (Beijing, Nov 2021)\cite{china_deploy_2021}\\
        Japan & \checkmark &  May 2020\cite{japan_level3_may2020} & Early 2022\cite{japan_level4} & With remote monitor\cite{japan_remote_2017} & - & Honda (Level 3 for congestion situation, Nov 2020)\cite{honda_2020}\\
        Korea & \checkmark & Jan 2020 \cite{korea_level3,korea_gov_level3} & - & - & - & - \\
        \hline\hline\\
    \end{tabular}
    \\
    * As of Jan 16, 2022, 32 of the 50 states in the US allow AV testing\cite{iihs_2022}\\
    ** As of Jan 16, 2022, 17 of the 50 states in the US allow deployment of AV, but to the authors' knowledge, details for the level of autonomy classification and due responsibility are not explicitly specified\cite{iihs_2022}\\
    *** As of Jan 16, 20 of the 32 states that allow AV testing do not have requirements of safety driver if certain conditions are met\cite{iihs_2022}\\
    **** Cruise applied for driverless deployment in California in May 2021\cite{cruise_driverless_app}, and received test permit in June 2021, then partial deployment approval in Sep 2021, waiting for final approval from California Public Utilities Commission\cite{cruise_driverless_permit}.
    \label{tab:legislation}
\end{table*}

\section{Certification}

\subsection{Certification Status Worldwide}
Since the piloting AV programs such as the PROMETHEUS project\cite{williams1988prometheus} (1987-1995) and DARPA Grand Challenge\cite{ozguner2007systems} (2004-2007), the development of automated vehicles have entered the era of on-road deployment. An incomplete summary of current certification of automated driving on public roads is provided in Table \ref{tab:legislation}.

\subsection{Safety Certification Challenges}

Among all the concerning issues, four particular challenges stand out for AV safety certification: the determination of unavoidable collision; the determination of liability; the verification cost for reasonable scenario coverage; and the additional cost of re-verification for ADS update.

Determining whether a collision situation is avoidable or not, this is not just an academic problem but also a liability-related legal problem. Despite the high expectations for autonomy, as humans we can only expect robotic systems to optimize actions for the \textbf{best possible} outcome. For example, one can only expect automated vehicles to apply evasive maneuver to avoid an avoidable collision, but not an unavoidable collision. The requirement for such avoidability determination has been passed as an amendment to law in Germany in August 2021 regarding level 3 to level 4 function in HAV\cite{schmid_2021}. Specifically, in \textsection 1e paragraph (2), number 2 of the law (Straßenverkehrsgesetzes, or Road Traffic Act) states that (English translation): 

\textit{Motor vehicles equipped with an automated driving function must have technical equipment that is capable to: independently comply with the traffic regulations directed at driving the vehicle and have an accident prevention system that:}
\begin{enumerate}[label=(\alph*)]
    \item \textit{is designed to avoid and reduce damage,}
    \item \textit{in the event of unavoidable alternative damage to different legal interests, the protection of human life being the highest priority, and}
    \item \textit{in the event of an unavoidable alternative risk to human life, none provides further weighting based on personal characteristics.}
\end{enumerate}

In the legal text above, judging whether ``unavoidable alternative damage" is the actual case requires a formal determination of damage avoidability, i.e. collision avoidability. Despite the different implementations of accident prevention systems in industry, from a safety verification standpoint, the avoidability judgement mechanism is better conservative than risky. At the same time, it is also undesirable to be ``too conservative" to not use emergency maneuver when an incoming collision is still avoidable. To illustrate this delicate difference, two situations are given:

\begin{enumerate}
    \item First, when the vehicle is very close to an unavoidable collision situation yet there is still room to avoid the collision, we don't want the ADS to make a false unavoidable judgement and start preparing for the impact;
    \item Second, when the vehicle is actually in an unavoidable collision situation and not colliding yet, from a safety verification standpoint, the ADS should be sensitive enough to immediately start preparing for the impact.
\end{enumerate}

  These two situations indicate that the accident prevention system in ADS should be able to make a clear-cut distinction between avoidable and unavoidable situations early enough, and it should not hesitate to engage conservative maneuver, evasive maneuver or preparation for impact. Otherwise, at least according to the German law amendment\cite{schmid_2021}, AV control is liable for collision damage.

\begin{figure}
    \centering
    \includegraphics[width = 0.48\textwidth]{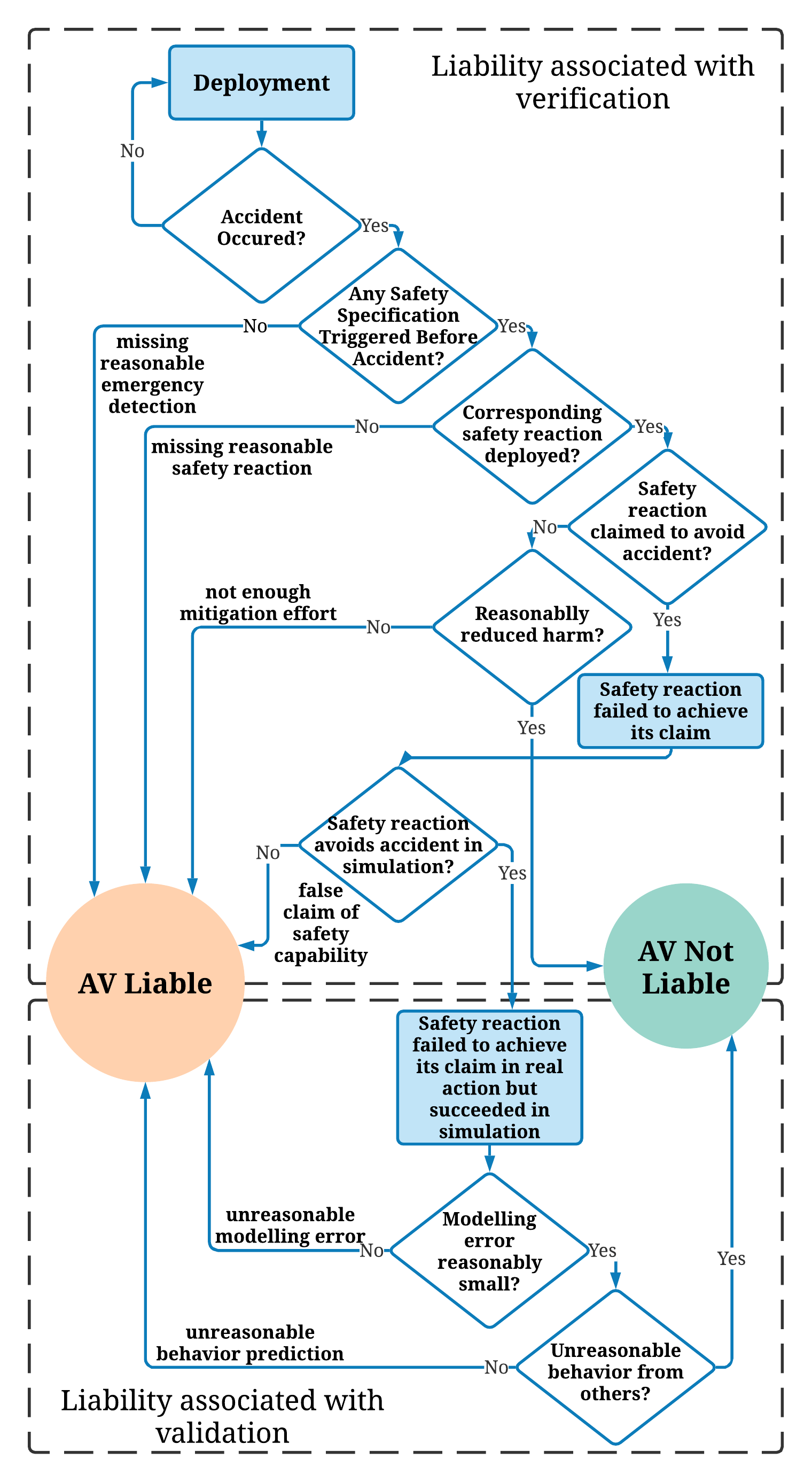}
    \caption{Proposed liability determination framework using formal method, for level 4 and level 5 ADS}
    \label{fig:liability_determine}
\end{figure}

Liability determination is also an emerging issue to be addressed for ADS deployment. At level 3 the ADS is not liable as long as unexpected or difficult situations are handed over early enough to the human driver, as the ADS is only expected to guarantee safety in its limited ODD environment. Even if the human driver does not take over in time, a fail-safe maneuver such as braking, stopping and turning on hazard lights can free the ADS from liability, to a certain extent. When a level 4 or above AV is equipped with formal safety verified ADS, the liability determination can be streamlined according to a formal logic, illustrated in Figure \ref{fig:liability_determine} for example. For ADS developed with sample-based safety verification, it is yet to see how a similar programmatic formal liability determination can be performed.

Another challenge for certification is the cost associated to verify and validate the ADS. For an ADS industry stakeholder, already huge investments have been made to verify and validate the in-house ADS. When it comes to type-approval or certification, it is not fully clear how much of the V\&V effort from the industry stakeholder can be ``credit transferred" towards the safety certification.

Moreover, as more learning-based functions are incorporated into AV control design, the resulting control stack is constantly evolving based on collected learning data. The continuous learning, together with the continuous development of ADS software by the engineering team is adding another layer of uncertainty to safety V\&V and certification. A question remains unanswered on whether regression testing is required for each major or minor ADS software update.

\section{Automated Vehicle Safety Verification and Validation}

\subsection{Qualitative Safety Goal}

Safety is system state free from harm. And safety assurance is the capability to be free from risks developing into harm when they emerge\cite{cohen2020constructive}. Although huge safety enhancement potential exists for AVs as 94\% of road accidents are driver-related\cite{national2015critical}, but limitations of current safety V\&V measures for AVs are widely recognized. As pointed out by RAND Corporation report \textit{Driving To Safety}\cite{kalra2016driving}, tens of billions of miles of on-road testing is required to ensure acceptable level of AV safety for all possible scenarios and inputs. Whereas programmatically generated simulation tests, although performed much faster, more flexible in scenario generation\cite{tahir2020coverage} and more economical, are susceptible to fidelity and realism concerns\cite{koopman2017autonomous}. Even worse, both on-road testing and simulated testing cannot exhaustively cover all driving scenarios even for a small operational design domain (ODD)\cite{J3016_202104}. 

The widely used safety guideline standard \textit{ISO 26262 Road Vehicles---Functional Safety}\cite{ISO26262} is suited only for mitigating known unreasonable risks related to known component failure (i.e. known unsafe scenario), but it is not addressing AV driving risks due to complicated environment variants and how the ADS reacts to them, while nothing in the vehicle is technically failing.

\begin{figure}
    \centering
    \includegraphics[width=.45\textwidth]{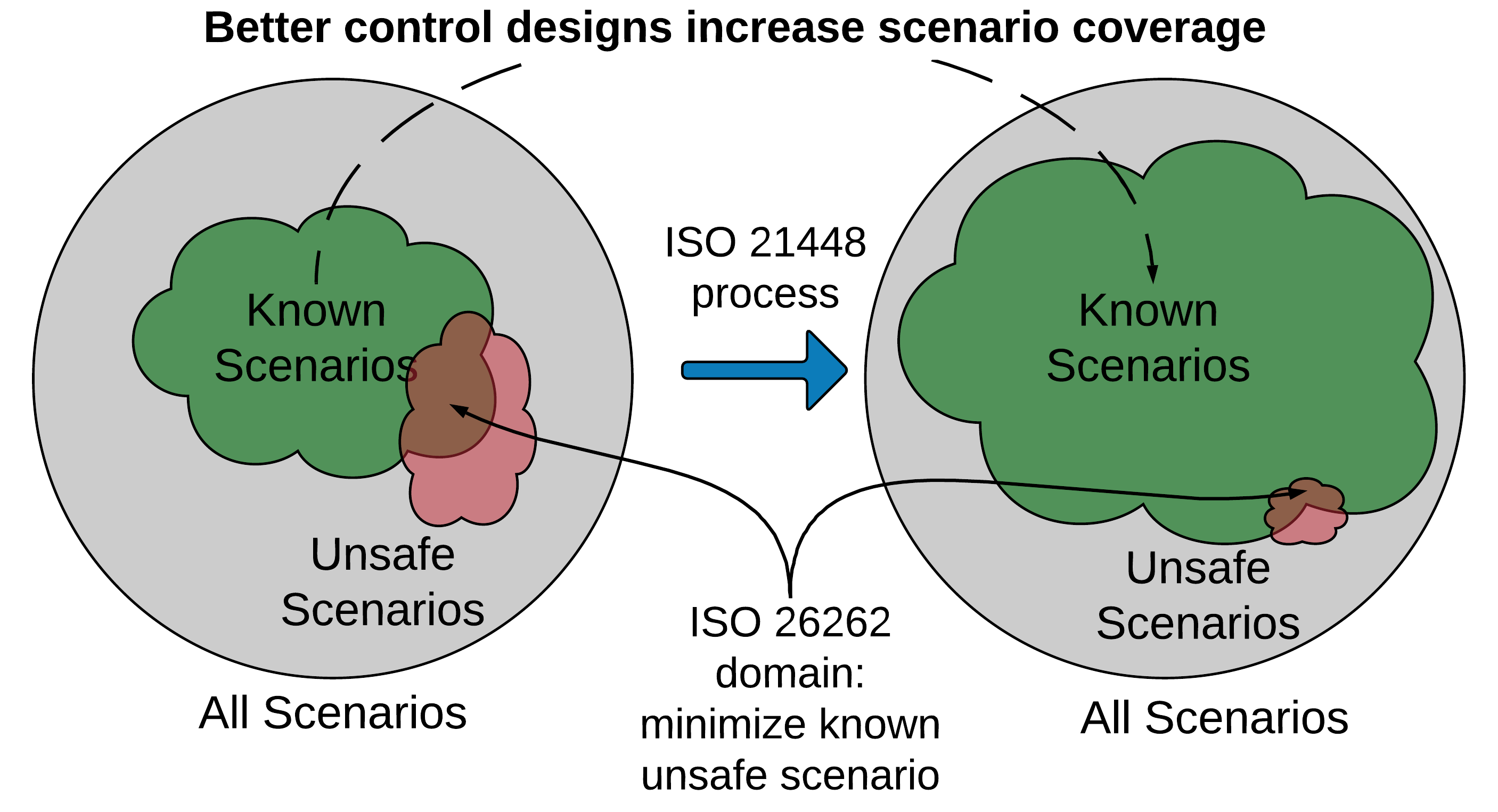}
    \caption{The goal of ISO 21448 (SOTIF)\cite{ISO21448}: minimize unsafe scenario outcomes, both known and unknown to the control design. Colors in the diagram represent: known safe scenarios (green), known unsafe scenarios (light red), unknown safe scenarios (grey), unknown unsafe scenarios (dark red).}
    \label{fig:SOTIF_goal}
\end{figure}

Given the safety challenge above, a qualitative goal was proposed in \textit{ISO 21448 Safety of the Intended Functions} (SOTIF)\cite{ISO21448}, which describes high-level goals of \textit{minimizing known and unknown unsafe scenario outcomes} for ADS function design (Figure \ref{fig:SOTIF_goal}). The challenge for meeting this goal, however, is the intractability of traditional methodologies such as field operational test (FOT)\cite{batsch2020taxonomy} to cover all possible scenarios for automated driving during testing.

Despite the challenge, there are promising methodologies and directions in safety analysis to move towards the ISO 21448 goal, such as maximizing scenario coverage in simulated testing\cite{tahir2020coverage}, or creating safety guarantees through formal methods\cite{chen2018signal,leung2020infusing}.

\subsection{Quantitative Safety Threshold}
For AV testing, large number of tested miles can be deceptive and does not always equal to adequate scenario coverage. Specifically, if most of the test miles are done on repeatedly tested, low traffic highway roads, the ``per mile value" of the test data quickly deteriorates. In fact, event rates per exposure notions such as injuries/fatalities per 100 million mile mentioned in the RAND report\cite{kalra2016driving}, or disengagements per 1000 miles used for California DMV disengagement report\cite{CA_DMV_disengagement} are only rough criteria to provide an intuitive summary of how statistically safe is a certain operation, with a risk of selective bias. 


In technical terms, \textit{safe enough} usually indicates full or sufficient scenario coverage for the designated ODD\cite{batsch2020taxonomy,riedmaier2020survey}. In reality, relatively weak requirements have been seen in existing regulations\cite{UN157} that only require to check ``a number of scenarios that are critical". A promising attempt to assign ``volume" to scenario samples is seen in \cite{weng2021} by introducing a ``$\delta$-neighborhood" around a discrete scenario. Mathematical algorithms such as T-wise\cite{grindal2005combination,amersbach_2019} and Poisson process\cite{aasljung2016comparing} have also been adopted to achieve ``almost full" statistical coverage by cleverly select candidates with limited samples. 

\begin{figure}
    \centering
    \includegraphics[width=0.48\textwidth]{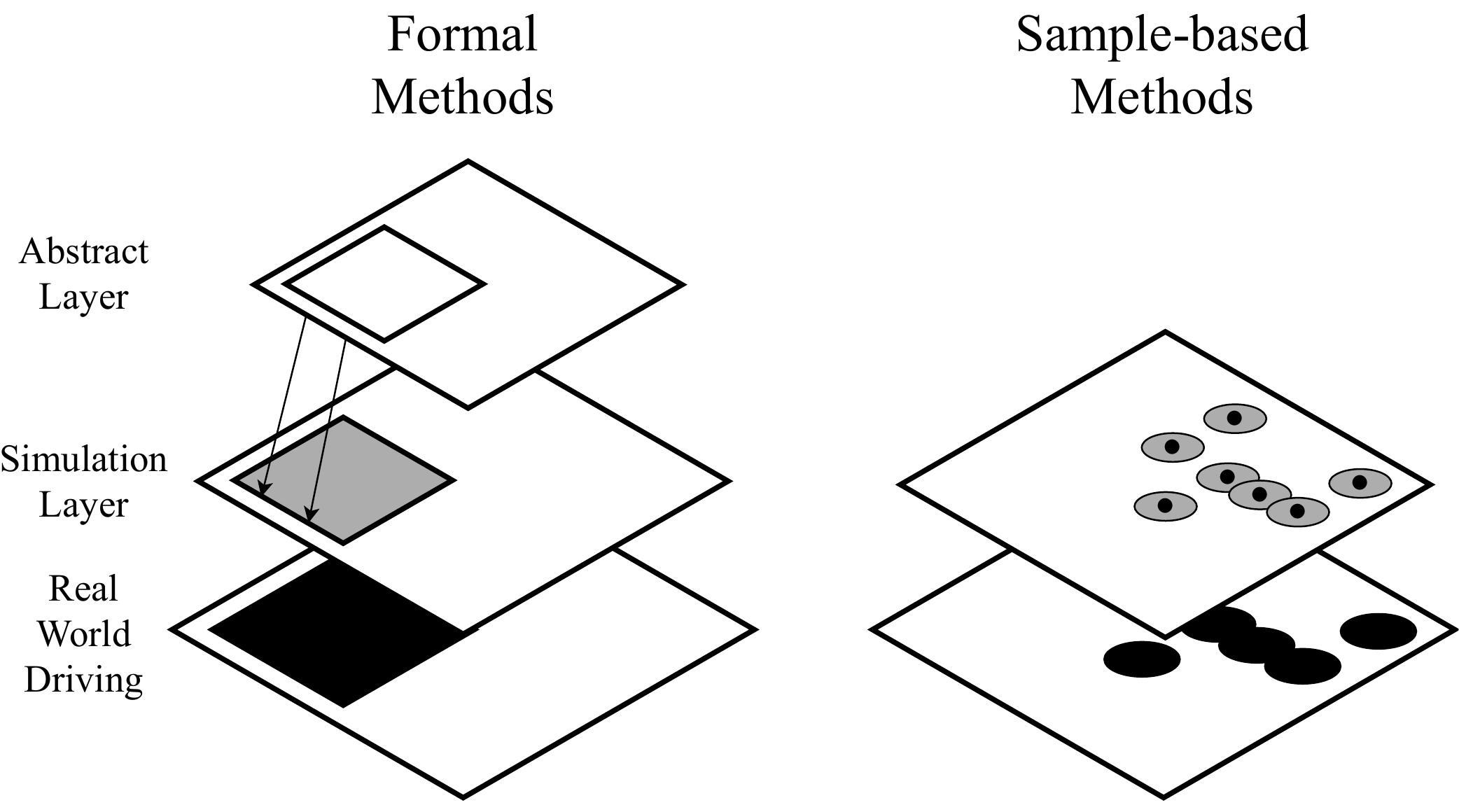}
    \caption{Scenario coverage comparison between formal methods and sample based methods: formal methods start from safety specification at a more abstract layer, may have larger single coverage volume in scenario space, but the procedure to integrate formal specification into control synthesis or monitoring can be controller math dependent and tedious; sample-based methods have more sporadic spread of scenario coverage due to the randomized generation process, but starts covering cases directly at the simulation layer, making the sampling process straightforward and easy to implement. Both approaches are trying to project maximum verified scenarios into real world driving, but the difference between simulation layer and real world driving always exists.}
    \label{fig:fm_sbm}
\end{figure}

Let us optimistically assume that this quantitative coverage-representation problem is solved and accepted across the community, and each single sampled scenario has a ``coverage volume" (illustrated in shaded areas in Figure \ref{fig:fm_sbm}, and in gridded units in Figure \ref{fig:evolution}(f)-(j)). Then the full scenario coverage task for sample-based methods is performing sufficient amount of sampled verification tests to ensure that each ``coverage volume" unit is associated with at least one safe test result. Realistically, some scenarios in a scenario space may be impossible to produce a safe test result (e.g. too little reaction distance with an obstacle), in this case, extra effort is required to confirm that those cases are indeed \textit{safety infeasible}. When it comes to type approval and certification, the authorities need to set an acceptable successful threshold for in order to match up with public expectations. The simplest threshold $r$ could be the ratio between the number of different tests verified safe $N_{\mathrm{sv}}$ and the number of minimum tests required for the ODD $N_{\mathrm{ODD}}$:
\begin{equation}
    r = \frac{N_{\mathrm{sv}}}{N_{\mathrm{ODD}}}
\end{equation}

In section \ref{sec:unified} we expand this scenario coverage concept to construct a unified framework of scenario coverage.



\subsection{State-of-the-art Verification and Validation Techniques}


Diversifying scenario sampling for testing is one of the major approaches for enhancing safe AV control in the development stage. For reference, entire surveys\cite{riedmaier2020survey,tahir2020coverage} have been dedicated to cover the various sub-categories of generating scenario samples for AV control testing. One of the main sub-categories, testing-based sampling, aims at maximizing scenario coverage at minimum effort; another sub-category, falsification-based sampling, aims at finding safety corner cases that are worth more attention for developers, such as high safety \textit{criticality} scenarios\cite{junietz_2018_criticality}. In terms of achieving the SOTIF goal of minimizing known and unknown unsafe scenarios, sample-based methods have less biased, more exploratory capability in discovering unknown unsafe scenarios, and the push from unknown to known is of ``horizontal" nature in the sense that all sampled scenarios are usually within a consistent simulation environment and the same fidelity level. While sample-based methods are not the focus of this review, it is worth mentioning that often a combination of methodologies\cite{riedmaier2020survey,kress2021formalizing,blumenthal2020safe}, both sample-based and formal methods, as well as field operational tests, is necessary during the development stage to boost the prototyping and design error discovery. In fact, novel approaches trying to combine sample-based methods with formal set-based methods have been proposed using a combination of resolution or probabilistic sample completeness and robust controlled invariance set\cite{weng2021}.

Compared to testing-based methods for safety verification and assurance, formal methods have the advantage of high statement reliability\cite{riedmaier2020survey} due to the rigorous logic foundation of formal methods. 

Commonly used formal methods in AV safety includes model checking, reachability analysis and theorem proving\cite{kress2021formalizing}. Model checking originates from software development to ensure that the software behavior is according to the design specifications. When safety specification is expressed in axioms and lemmas as in\cite{shalev2017formal}, then theorem proving can be used to verify safety using worst case assumptions. Reachability analysis\cite{antsaklis1997linear} has a special place among the three due to its inherent capability of producing safety statements for dynamical systems\cite{bertsekas1971minimax}, which capture the main characteristic of dynamic driving task (DDT).

\begin{figure}[t]
    \centering
    \includegraphics[width=0.45\textwidth]{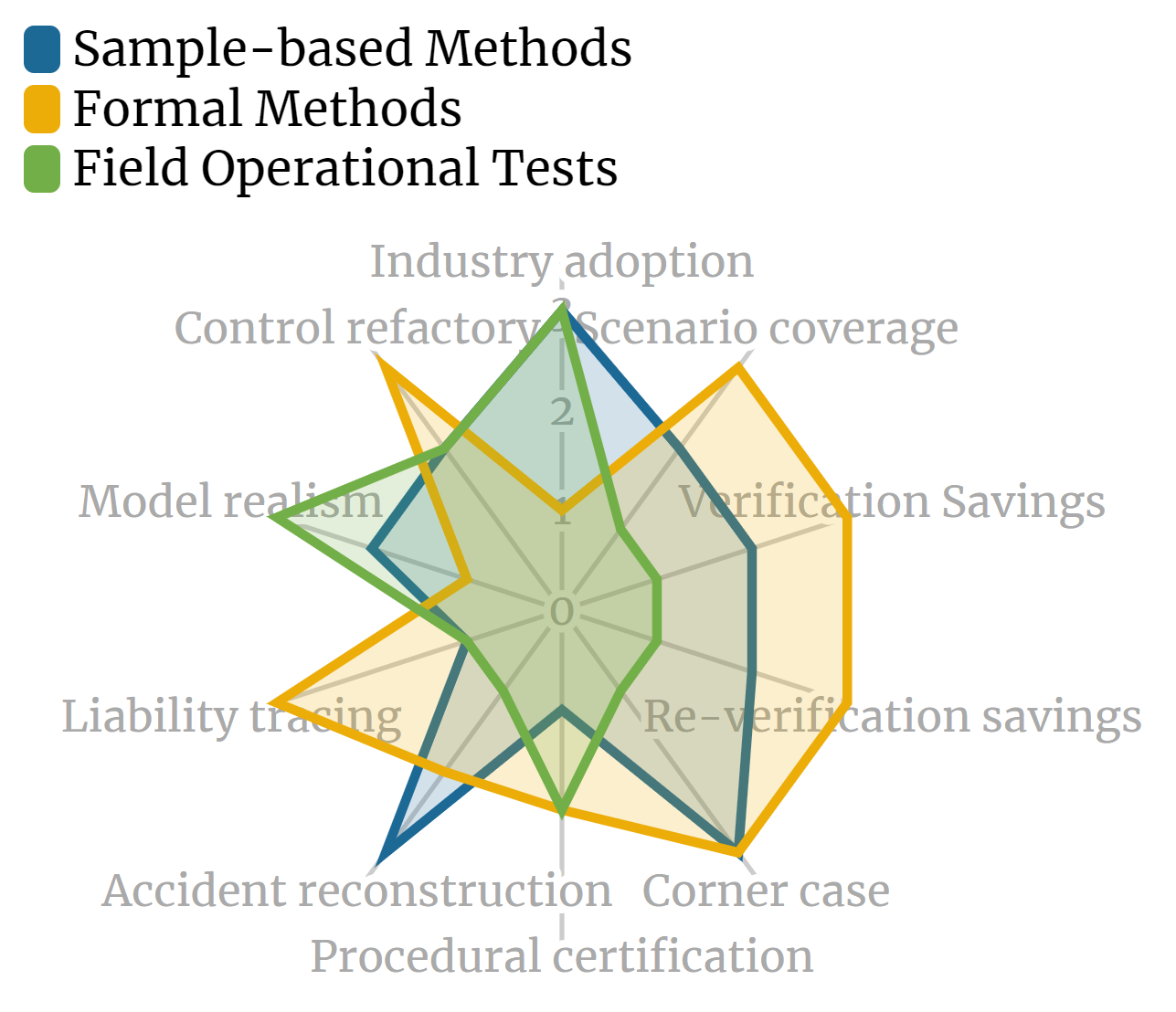}
    \caption{Comparison of different verification and validation methods. \textbf{Industry adoption}: sample-based methods have been widely used to generate virtual scenarios to complement limited road tests, while formal methods are less common. \textbf{Scenario coverage}: in a properly abstracted model formal methods can cover the entire scenario space by giving explicit specifications, while sample-based methods require large numbers of samples to establish coverage, and road tests can hardly cover corner cases. \textbf{Verification savings}: Relying on proofs and dynamics, formal methods can save the most in verification cost, while road tests are more expensive than sample-based virtual test due to hardware and personnel costs. \textbf{Re-verification savings}: formal methods save the most for re-verification since often minor update in the proof is needed; the need for regression testing after control update makes sample-based and especially road tests expensive. \textbf{Corner case}: corner case generation is well studied for sample-based methods, formal methods do not explicitly focus on corner cases but covers them; for road tests corner cases are difficult to encounter. \textbf{Procedural certification}: formal methods have the potential to fit with certification with its rigor in theoretical proof, while road tests are practically implementable but coverage-wise insufficient, while sample-based testing lacks certifiable validity. \textbf{Accident reconstruction}: can be easily performed using sample-based scenario generation, but not economical to be reproduced with real vehicle tests; formal methods can utilize accident information to recheck the execution of formal safety specifications for design flaws. \textbf{Liability tracing}: can be performed naturally with formal methods as shown in Fig. \ref{fig:liability_determine}. \textbf{Model realism}: is well matched in road tests, less matched in sampling-based simulation tests, and least matched in models used for formal methods due to the known issue of scalability. \textbf{Control refactory}: can be relatively easy with formal methods based verification, as long as new designs go through the same formal verification pipeline, whereas in sample-based testing or road tests new extra efforts are unavoidable to re-verify the changes.}
    \label{fig:vv_compare}
\end{figure}

Real world road testing, or field operational testing (FOT), is the ultimate yet expensive approach for AV verification and validation. In a certain sense it is the only way of validation. Yet the shortcomings for FOT is also obvious: it lacks the ability to reach enough scenario coverage (especially near collision and collision scenarios) with a candidate AV controller on board. A comparison of sample-based, formal and field operational testing methods is illustrated in Figure \ref{fig:vv_compare}. The different aspects are scored on a scale of 0 to 10, with 10 representing highest satisfaction.

\section{Formal Safety Verification}
Formal methods are a class of methods that applies mathematically rigorous techniques (usually in the form of logic calculi) to achieve specification and verification of software and engineering design\cite{wing1990specifier}. By definition it has inherent advantage for safety verification tasks, but highly system complexity with continuous dynamics has been a major limiting factor for its wide application in automated driving, i.e. suffering from scalability\cite{koopman2017autonomous,kapinski2016simulation}. Technically, formal methods can be generalized as the process of implementing or translating abstract specifications into system control algorithm or program, such that the controlled system behavior will satisfy the said specifications. 

\subsection{Preliminary: Reachability Analysis}

We first define a dynamical system with continuous states and controls:
\begin{equation}\label{eq:system}
    \dot{x} = f(x,u)
\end{equation}
where state $x\in \mathcal{X} \subseteq \mathbb{R}^n$ and control $u\in\mathcal{U}\subseteq \mathbb{R}^m$ are measurable and continuous within their domains, $\mathcal{X},\mathcal{U}$ are compact, $m,n\in\mathbb{R}$. The function (i.e. flow field) $f: \mathbb{R}^n\times \mathbb{R}^m \xrightarrow{} \mathbb{R}^n$ is assumed to be uniformly continuous, bounded and Lipschitz continuous.

\textit{Definition (Maximal Forward Reachable Set)}: given a dynamical system as in (\ref{eq:system}), the forward reachable set at time $t>t_0$ of a set of initial states $\mathcal{X}_{\mathrm{init}}(t_0)\subseteq \mathcal{X}$ is defined as:
\begin{equation}
\begin{aligned}
    maxFRS:\quad \{x(t)\in\mathcal{X} ~|~ \exists u\in\mathcal{U}, \exists x(t_0)\in \mathcal{X}_{\mathrm{init}}(t_0), \\ x(t)= x(t_0)+ \int_{t_0}^t f(u,t) \mathrm{d}t \}
\end{aligned}
\end{equation}

Forward reachable set propagates dynamics to find all possible reachable states in a future time $t$ since current time $t_0$ from within $\mathcal{X}_{\mathrm{init}}$. On the contrary, backward reachable set back-propagates to find all possible states from a previous time $t$ that can lead to some target state set $\mathcal{X}_{\mathrm{goal}}$ at current time $t_0$:

\textit{Definition (Maximal Backward Reachable Set)}: given a dynamical system as in (\ref{eq:system}), the backward reachable set at time $t< t_0$ of a set of target states $\mathcal{X}_{\mathrm{goal}}(t_0)\subseteq \mathcal{X}$  is defined as:
\begin{equation}
\begin{aligned}
    maxBRS:\quad \{x(t)\in\mathcal{X} ~|~ \exists u\in\mathcal{U}, \exists x(t_0)\in \mathcal{X}_{\mathrm{goal}}(t_0), \\ x(t_0)= x(t)+ \int_{t}^{t_0} f(u,t) \mathrm{d}t\}
\end{aligned}
\end{equation}

The operator $\exists$ before $u$ in the above expressions can be replaced with $\forall$ to create the ``minimal" reachable sets, for example, minimal backward reachable set:
\begin{equation}
\begin{aligned}
    minBRS:\quad \{x(t)\in\mathcal{X} ~|~ \forall u\in\mathcal{U}, \exists x(t_0)\in \mathcal{X}_{\mathrm{goal}}(t_0), \\ x(t_0)= x(t)+ \int_{t}^{t_0} f(u,t) \mathrm{d}t\}
\end{aligned}
\end{equation}

In some cases, reachability within defined time horizon is of more interest. Thus the definitions in the previous section can be extended to include time from current time to end of time horizon. For brevity, only the definition of maximal forward reachable tube is provided, the other variants can be easily derived accordingly:

\textit{Definition (Maximal Forward Reachable Tube)}: given a dynamical system as in (\ref{eq:system}), the forward reachable tube during time $[t_0,t]$ of a set of initial states $\mathcal{X}_{\mathrm{init}}(t_0)\subseteq \mathcal{X}$ is defined as:
\begin{equation}
\begin{aligned}
    maxFRT:\quad \\
    \{x(s)\in\mathcal{X} ~|~ \exists u\in\mathcal{U}, \exists s\in [t_0,t], \exists x(t_0)\in \mathcal{X}_{\mathrm{init}}(t_0), \\ x(s)= x(t_0)+ \int_{t_0}^s f(u,t) \mathrm{d}t \}
\end{aligned}
\end{equation}

In situations where the interaction among different traffic participants play a critical role, it becomes imperative to include the influence of participating, sometimes even adversarial agents into the reachability analysis. In this case, an additional input $d$ is introduced to the system (\ref{eq:system}) to represent such adversarial control input:
\begin{equation}\label{eq:system_adversarial}
    \dot{x} = f_{a}(x,u,d)
\end{equation}
where $d\in\mathcal{D}\subseteq \mathbb{R}^p, p\in\mathbb{R}$ is the adversarial input to the system. As an example, it can be control decisions of other vehicles where $f_a: \mathbb{R}^n\times \mathbb{R}^m \times \mathbb{R}^p\xrightarrow{} \mathbb{R}^n$ represents the joint or relative dynamics of all participating vehicles. In such systems, the influence of adversarial inputs is not up to the ego vehicle control, and therefore must be modelled conservatively to provide guarantees in worst cases. The definition of maxFRS and minBRT in such systems are provided below as examples.

\textit{Definition (Maximal Forward Reachable Set under Adversarial Influence)}: given a dynamical system as in (\ref{eq:system_adversarial}), the forward reachable set at time $t>t_0$ of a set of initial states $\mathcal{X}_{\mathrm{init}}(t_0)\subseteq \mathcal{X}$ is defined as:
\begin{equation}
\begin{aligned}
    \{x(t)\in\mathcal{X} ~|~ \exists u\in\mathcal{U}, \forall d\in\mathcal{D}, \exists x(t_0)\in \mathcal{X}_{\mathrm{init}}(t_0), \\ x(t)= x(t_0)+ \int_{t_0}^t f(x,u,d) \mathrm{d}t \}
\end{aligned}
\end{equation}

While the above definitions have been made in a deterministic sense, reachability can be stochastically interpreted, e.g. by determining the probability of system states reaching a certain target set\cite{prandini2006stochastic,abate2008probabilistic}.

\subsection{Signal Temporal Logic}
In the field of computer science and robotics, Signal temporal logic (STL) is a common language versatile for expressing and specifying requirements that is time critical and containing continuous variables. The basic semantic of signal temporal logic is listed in table \ref{tab:STL}, adapted from \cite{chen2018signal,arechiga2019specifying}. In brief, STL uses first order logic\footnote{First order logic: a collection of formal systems that uses quantifiables over non-logic objects such as variables. For example, while propositional logic can only make statements like ``the vehicle violates lane boundaries", first order logic can incorporate variables and state ``there exists a coordinate $C$ such that $C$ is inside the vehicle occupancy area and $C$ is outside the intended lane boundaries $B$". In the later statement, ``exists", ``inside" are quantifiables, and $C$ is a variable.} to make statements about variables in their temporal development.

\begin{table*}[ht]
    \centering
    \begin{tabular}{p{3cm}p{1cm}p{4cm}p{8cm}}
        \hline\hline\\
        STL semantic* &  & Equivalent STL semantic & Explanation\\
        \hline\\
        $(s,t)\models f(\cdot) \geq 0$  & $\Longleftrightarrow$ & $f(s(t)) \geq 0$ & $f(s(t)) \geq 0$ holds true for the trajectory $(s,t)$ at time $t$ \\
        $(s,t)\models \neg(\phi)$ & $\Longleftrightarrow$ & $\neg((s,t)\models\phi)$ & The negation of predicate\footnote{Predicate: in mathematical logic theory, predicate is a function that returns either true or false.} $\phi$ holds for the trajectory $(s,t)$ at time $t$\\
        $(s,t)\models \phi \land \psi$ & $\Longleftrightarrow$ & $(s,t)\models \phi \land (s,t)\models \psi$ & Both predicates $\phi$ and $\psi$ hold simultaneously for trajectory $(s,t)$ at time $t$\\
        $(s,t)\models \phi \lor \psi$ & $\Longleftrightarrow$ & $(s,t)\models \phi \lor (s,t)\models \psi$ & Either predicate $\phi$ or $\psi$ holds for trajectory $(s,t)$ at time $t$\\
        $(s,t)\models \square_I \phi$ & $\Longleftrightarrow$ & $\forall t'\in I\bigoplus t \quad s.t.\quad (s,t')\models \phi$ & Predicate $\phi$ holds for the entire interval $I$\\
        $(s,t)\models \lozenge_I \phi$ & $\Longleftrightarrow$ & $\exists t'\in I\bigoplus t \quad s.t.\quad (s,t')\models \phi$ & Predicate $\phi$ holds at least once in the interval $I$\\
        $(s,t)\models \phi\mathcal{U}_I \psi$ & $\Longleftrightarrow$ & $\exists t'\in I\bigoplus t \quad s.t.\quad ((s,t')\models \square_{[0,t']} \phi) \land ((s,t')\models \psi)$ & Predicate $\phi$ always holds until predicate $\psi$ is true at some moment $t'$ in interval $I$. After moment $t'$ predicate $\phi$ or $\psi$ no longer has to hold\\
        
        \\
        \hline\hline\\
        \multicolumn{4}{l}{*Note: when $I$ is dropped in the subscript (as is most cases), it indicates that $I=[0,\infty)$ by default.}
        
    \end{tabular}
    \caption{Basic STL semantics}
    \label{tab:STL}
\end{table*}

An exemplary casual safety specification for AVs can be ``never cause a collision in traffic", but once translated into an STL specification, some of the ambiguities in the casual specification needs to be removed. First, the word ``cause" is not well defined in STL as it involves the complexity of collision liability determination, and may have to be replaced with the phrase ``be in", which is liability-neutral. Then the casual specification becomes ``never be in a collision in traffic". Second, STL requires a precisely defined scope of time for the specification, and for automated driving it is a common practice to use the concept of time horizon to narrow down the time span to a practical and tractable level. Therefore the casual specification can be further modified into ``never (in the time horizon) be in a collision". At this point, the modified casual safety specification can be translated into a simple STL formula:
\begin{equation}
    (s,t)\models \square_I \phi^{\text{collision-free}}
\end{equation}
This express means that during the time horizon interval $I$ the trajectory of the automated vehicle system always satisfies the requirement $\phi^{\text{collision-free}}$ (the automated vehicle state is in the collision free set $\mathcal{S}^{\text{collision-free}}$).

The above STL-translated safety specification still stays at an abstract level in set language, and it is the responsibility of AV control algorithm designers to honor this safety specification by either synthesizing it into the controller architecture, or performing safety verification over the designed controller to ensure that the specification is met with enough confidence or solid proof. Some examples of STL-based safety specifications are: (1) always keep the ability to return to the right-most lane within 5 seconds; (2) maintain collision-free and within road boundaries. When a finite time horizon (such as $I= [0,25](s)$) is provided, the specifications (1) and (2) can be expressed as\cite{chen2018signal}:
\begin{equation}
    (s,t) \models \square_{[0,25]}\lozenge_{[0,5]}\psi^{lane} \land \square_{[0,25]}\phi
\end{equation}
where $\psi^{lane}$ represents the specification of being always within 5-seconds of returning to the right lane, and $\phi$ represents collision-free and within road boundaries. The expression $\square_{[0,25]}\lozenge_{[0,5]}\psi^{lane}$ encodes ``always" ($\square_{[0,25]}$) ``have the existing capability to satisfy in 5 seconds" ($\lozenge_{[0,5]}$) the ``going back to right lane" requirement ($\psi^{lane}$). The expression $\square_{[0,25]}\phi$ encodes ``always" ($\square_{[0,25]}$) ``satisfy collision-free and road boundary constraints" ($\phi$).  More sophisticated control goal can be encoded based on this STL expression, for example, to indicate the task of overtaking a slow car in front (expression modified from \cite{chen2018signal}):
\begin{equation}\label{eq:stl_example}
    (s,t) \models \square_{[0,25]}\lozenge_{[0,5]}\psi^{lane} \land ( \phi \mathcal{U}_{[0,25]}\psi^{pass} \lor \phi \mathcal{U}_{[0,25]}\psi^{stay} )
\end{equation}
where $\psi^{pass}$ represents successful overtaking the slow front car, and $\psi^{stay}$ represents staying behind the slow front car. The expression $( \phi \mathcal{U}_{[0,25]}\psi^{pass} \lor \phi \mathcal{U}_{[0,25]}\psi^{stay} )$ specifies two alternative goals ($\psi^{pass}$ or $\psi^{stay}$) on top of the safety constraint $\phi$. STL expressions like equation (\ref{eq:stl_example}) were implemented into control synthesis via connection between STL and reachability analysis in some recent work\cite{chen2018signal}. Apart from control synthesis, safety specifications expressed in STL can be used as assertions\footnote{Assertion: in computer science, an assertion is a predicate connected to a point in the program that should always evaluate to true if the program is implemented correctly.} during controller prototyping to indicate safety violations, so that developers are constantly aware of safety specifications during control design. The logic calculation for determining the truth of a safety specification is often done by solving a satisfiability modulo theory (SMT) problem\footnote{Satisfiability modulo theory (SMT) problem: decision problem that returns true or false for formulae expressed using first order logic, for reference see \cite{barrett2018satisfiability,de2011satisfiability,bjorner2012program,barrett2005smt}.}.

\subsection{The Formal Safety Philosophy}

Unlike the sampling-based methods where the verification against scenario variations is done through populating large number of scenario samples, in formal verification, the verification is mostly done when the safety specification is implemented in the controller to the fidelity level of the simulated environment. This is due to the different mechanism of how safety is verified. In the formal safety philosophy, a safety specification is either fulfilled or violated, and the fulfillment property can be \textbf{designed} by synthesizing the specification fulfillment into model-based control design. After such design is complete, the burden of ensuring such fulfillment during operation is transferred to online validation of model correctness: as long as the control-oriented model is validated to be correct, and the controller executes by the synthesized safety specification, then the system is provably safe. Without loss of generality, a safety specification $\phi$ may not be feasible in certain situations (such as an unavoidable collision), $\forall u, (s,t) \not\models \phi$. In this case, the $\phi$-synthesized controller cannot find a feasible control sequence to fulfill $\phi$, and the best practice would be to promote the situation to premeditated emergency or fall-back strategy (such as collision impact preparation). In any event, since the $\phi$-synthesized controller by design has exhausted its available actions and still cannot find a $\phi$-fulfilling action, the controller is not at-fault for the unavoidable damage.


\section{Unified Scenario Coverage Framework}\label{sec:unified}

\subsection{Scenario Space and Scenario Volume}

\begin{figure}[ht]
    \centering
    \includegraphics[width=.48\textwidth]{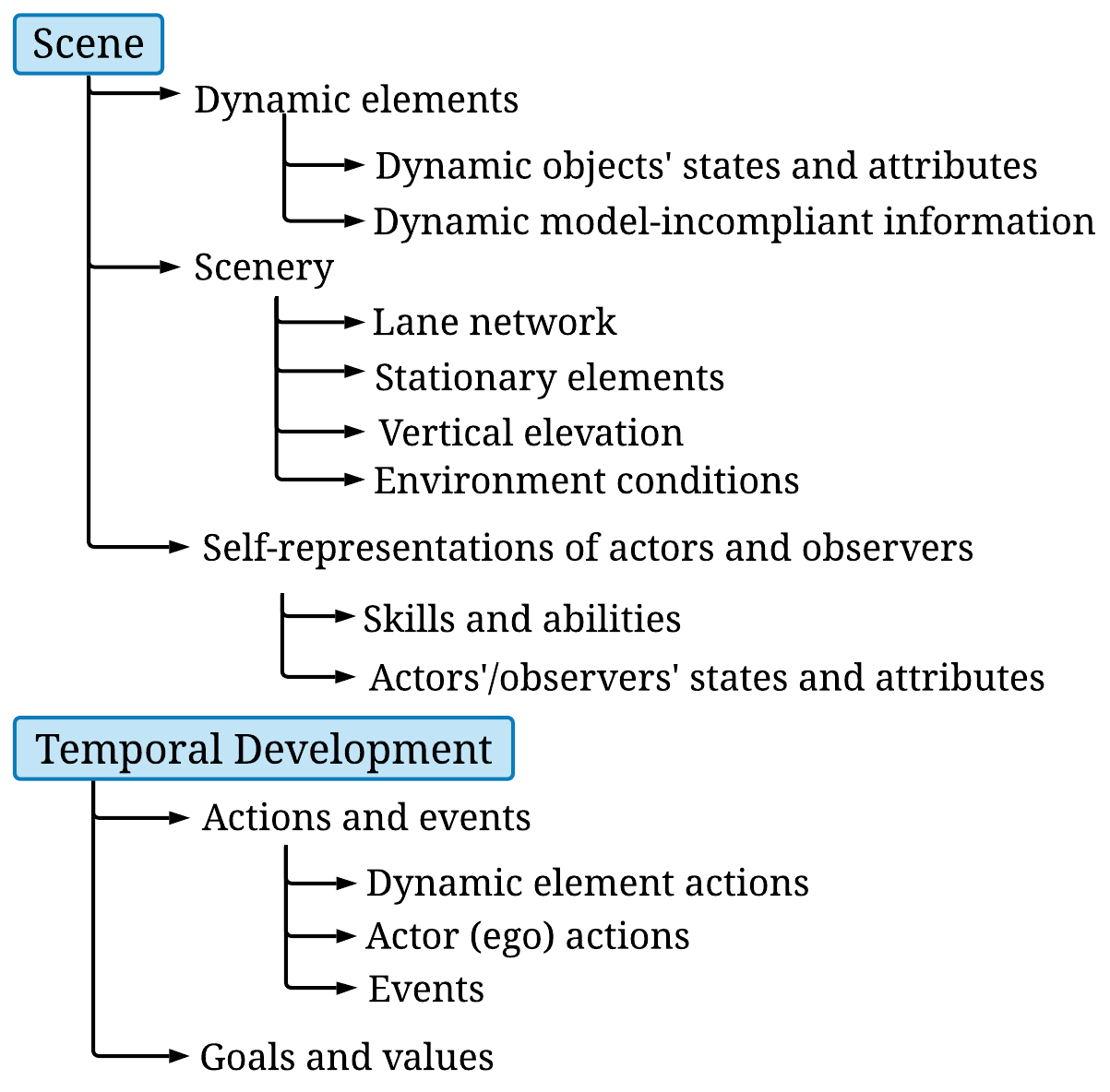}
    \caption{Dimensions of scenario variation: according to ISO 21448\cite{ISO21448}, the initial scene associated with a scenario can have multiple variations including dynamics elements, scenery setting and self-representations of actors and observers; given an initial scene, the temporal development can have additional variations of dynamic element actions, actor (ego) actions and goals and values of different participants in the scenario.}
    \label{fig:scenario_variation}
\end{figure}

According to ISO 21448\cite{ISO21448}, \textit{scenario} is a description of the temporal development between several scenes in a sequence of scenes, and \textit{scene} is a snapshot of the environment including the scenery, dynamic elements, and all actor and observer self-representations, as well as the relationship among those entities. Complying to this definition, the task of \textit{ensuring full safety} in a defined ODD can be described as performing safety verification and validation for all possible scenario variations within the ODD. The variations of scenario within an ODD is two-fold: variations of the initial scene, and variations of the temporal development since the initial scene. The different dimensions of scenario variation is illustrated in Figure \ref{fig:scenario_variation}.

Given the specific ODD designated as $\mathcal{O}$, one can define an entity to span all possible scenarios within the ODD description. We call this the \textit{scenario space}. Using parameterization, the \textit{scenario space} is defined as:

\textbf{Definition 1. (Scenario space)} The N-dimensional sub-space $S \subset \mathbb{R}^\mathrm{N}$ where $S = \{(p_1,\cdots,p_\mathrm{N},q_1,\cdots,q_\mathrm{M})\in \mathbb{R}^\mathrm{N+M}| p_i \in [\underline{p_i},\overline{p_i}], q_j\in \mathcal{Q}_j \}$ is the scenario space for an ODD specified by the continuous scenario parameters $p_i (i= 1,\cdots,\mathrm{N})$ and discrete scenario parameters $q_j (j=1,\cdots,\mathrm{M})$. $\underline{p_i},\overline{p_i}\in\mathbb{R}$ are the lower and upper bounds of each continuous parameter $p_i$, $\mathcal{Q}_j \subset \mathbb{R}$ is the set of possible values for each discrete parameter $q_j\in \mathcal{Q}_j$.

\textbf{Remark 1.} \textit{Note that in such a parameterized definition, no description of stochastic behavior is explicitly given, since the parameters are used to capture the scenario event as if it has already happened and fully describable without stochasticity.}

Assume that the number of continuous dimensions of scenario space $S$ is $\mathrm{N}$, the number of discrete parameter combinations is $\prod_{j=1,\cdots,\mathrm{M}}|\mathcal{Q}_j|$, where $|\mathcal{Q}_j|$ is the cardinal of parameter set $\mathcal{Q}_j$. Assume the feasible range of each continuous parameter $p_i$ is $[\underline{p_i},\overline{p_i}]$, then the total \textit{scenario space volume} can be defined as:

\textbf{Definition 2. (Scenario space volume)} The scalar value $V_{S(\mathcal{O})}=\prod_{i=1,\cdots,\mathrm{N}} (\overline{p_i}-\underline{p_i})\cdot \prod_{j=1,\cdots,\mathrm{M}}|\mathcal{Q}_j|$ is the scenario space volume of the scenario space $S(\mathcal{O})$ specified by the ODD $\mathcal{O}$. The meaning of symbols are aligned with Definition 1.


\textit{Full scenario coverage} for safety verification requires that the verified scenario space volume $V_v$ encompasses the volume $V_S(\mathcal{O})$:
\begin{equation}
    V_v \supseteq V_{S(\mathcal{O})} =\prod_{i=1}^{\mathrm{N}} (\overline{p_i}-\underline{p_i})\cdot \prod_{j=1}^{\mathrm{M}} |\mathcal{Q}_j|
\end{equation}
where $V_v$ is safety verified scenario volume.

Similarly, \textit{partial scenario coverage} indicates that safety verification is carried out only on a portion $V_v{'}$ of the scenario space volume $V_{S(\mathcal{O})}$:
\begin{equation}
    \left(V_v{'} \cap V_{S(\mathcal{O})} \right) \subset V_{S(\mathcal{O})} 
\end{equation}

The scenario coverage ratio is therefore:
\begin{equation}
    \frac{V_v{'} \cap V_{S(\mathcal{O})}}{V_{S(\mathcal{O})}}
\end{equation}

\begin{figure*}[t]
    \centering
    \includegraphics[width=.95\textwidth]{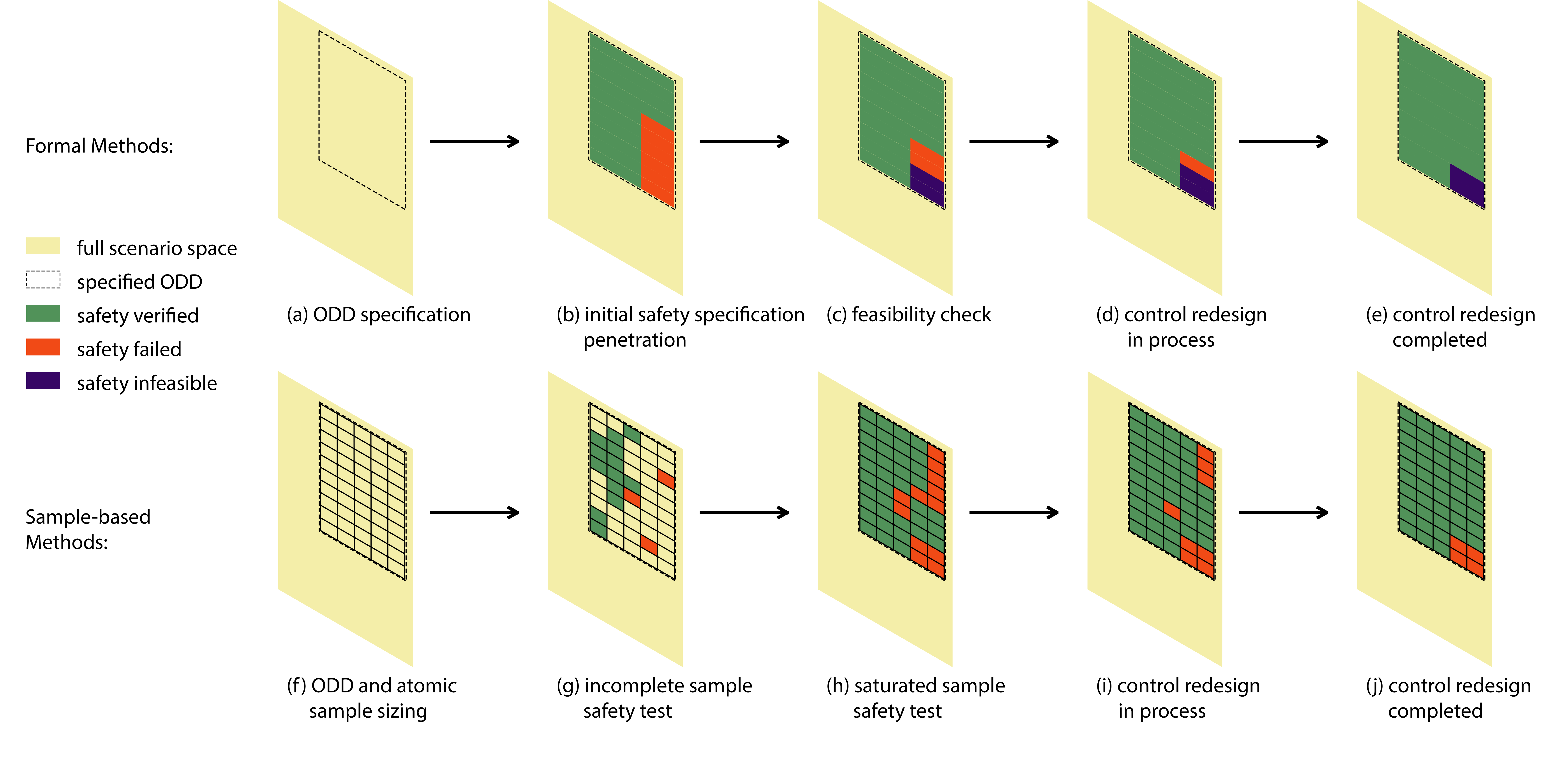}
    \caption{Control policy evolution for safety verification. In (a)-(e), formal methods start from safety specification for the ODD (a), then an initial safety specification penetration test is performed to see how good the safety specification is kept (b). Then the feasibility check is performed to see how much of the failed scenarios are actually safety feasible (c). Then the AV controller is redesigned to remedy the failed scenario volume region(d), and eventually all safety feasible scenario volume is verified safe with the candidate AV controller. In (f)-(j), the sample-based methods start by segmenting the ODD into verifiable scenario units (f), then an incomplete sampling safety test is performed to check for major issues with the candidate AV controller (g). Subsequently fully saturated sampling is performed to ensure scenario coverage (h). After full scenario coverage test, controller weakness is exposed and the redesign process iterates (i). Eventually the controller weakness stops to shrink, and the control redesign process is likely finished(j). Notice that with formal methods safety infeasible scenario regions can be identified, and no further redesign is needed once all volume in the ODD except for the safety infeasible scenario volume is verified.}
    \label{fig:evolution}
\end{figure*}

\subsection{Unified Scenario Coverage}
Sample-based methods do not explicitly possess the volume property associated with each sample scenario, as each scenario sample is rather atomic and volume-less. In order to make unified comparison and utilization of both methods for safety verification, notions of scenario volume must be assigned to sample-based methods. For simplicity, assume all $N$ continuous dimensions are orthogonal to each other, and a axis-aligning polygonal volume in $N$-dimensional scenario space is assigned for a scenario sample specified by the parameter tuple $\left(p_1^0,p_2^0,\cdots,p_N^0\right)$. The assigned volume will take the width of $2w_{i}$ in the $i^\mathrm{th}$ continuous scenario space dimension around the parameter value $p_i^0$, where $w_i$ is the tolerable resolution provided by the verification requirement:

\textbf{Definition 3. (Unit scenario volume)} Given a scenario space with $N$ continuous dimensions and tolerable resolution of $w_i, (i=1,\cdots,N)$ in each dimension, the unit scenario volume is:
\begin{equation}
    V_{0} =\prod_{i=1,\cdots,N} (2w_{p_i})
\end{equation}

Thus verifying full scenario coverage with sample-based methods requires to generate enough distinct, evenly spaced samples such that the overlapped volume of all verified samples supersets $V_{S(\mathcal{O})}$. The number of samples required is:
\begin{equation}
    n= \lceil \frac{V_{S(\mathcal{O})}}{V_0} \rceil
\end{equation}
where $\lceil \cdot \rceil$ is the ceiling function.

\textbf{Remark 2.} \textit{The number of samples required $n$ can possibly be reduced if evidence can be provided to show that verifying certain scenario sample can simultaneously lead to the verification of a certain group of scenario samples.}

We also define safety for this unified framework. Without loss of generality, safety can be defined as a specification in natural language:

\textbf{Definition 4. (Safety Specification)} Safety is the verifiable truthfulness of statement specified by one or more safety requirement clause(s) $\mathcal{N}$ in human natural language.

As an example, $\mathcal{N}$ could be \textit{``the automated vehicle should be collision-free at all times"}, or \textit{``the automated vehicle should be fault-free at all times"}. If translated to STL, these two safety specifications can be expressed as:
\begin{equation}
    (s,t)\models \square \phi^{\mathrm{collision-free}},\quad (s,t)\models \square \phi^{\mathrm{fault-free}}
\end{equation}

In sample-based methods, the safety of a simulated scenario is checked at the end of each scenario sample. On the other hand, formal methods usually start with the safety specification before any simulation is performed, then the specification is translated into machine-interpretable language statement, for example, in linear temporal logic (LTL) or signal temporal logic (STL). Then the translated specification can be utilized by checking the validity of the specification in simulated/real world tests or by turning the specification into system constraints or other control design functions during control synthesis. 

Ideally, for formal methods, if all safety related specifications are truthfully translated into such machine-interpretable statement in the first place, and if the synthesized controller fully respect these statements with a model-based controller that perfectly models the (simulated) environment, then the scenario coverage is 100\% with respect to the simulation layer. 

\textbf{Definition 5. (Specification translation)} is the process of turning a specification $\mathcal{N}$ in definition 3 to an interpreted equivalent verifiable expression $E_{\mathcal{N}}$ expressed in verification model states $(s,t)$.

As an example, the expression \textit{``the automated vehicle should be collision-free at all times"} can be translated into a verifiable expression in the form of a mathematical function that evaluates occupancy overlap between the automated vehicle and any other objects in the simulation model and return true if overlap happens in the simulation. When a control policy is specified, we have the following variant:

\textbf{Definition 6. (Policy controlled specification translation)} is a specification translation under the influence of a control policy $\pi$ to the system.

However, practical challenges prevent ideal 100\% translation of formal safety specifications due to the following reasons: first, formal methods rely on arguments based on abstraction of the real world, therefore discrepancies -- whether large or small -- between abstraction and reality, will occur and impair the effective safety guarantee. Second, the actual developed controller usually have performance limitations that prevents compliance with the safety specification in every case. Third, practical scalability difficulties exist in translating formal safety specifications into safety verification or control synthesis. In other words, formal methods usually have a \textit{penetration rate} less than 100\% in reality. Formally we define the penetration rate as: \textit{the percentage of verifiable scenarios in the scenario subspace specified by a formal safety specification}:

\textbf{Definition 7. (Specification penetration rate)} The ratio between specification holding volume $V_h$ and specification claimed scenario space volume $V_{S(\mathcal{N,\pi})}$ during a policy $\pi$-controlled specification translation is the specification penetration rate $\mathcal{PR}(V_{S(\mathcal{N},\pi)},\pi)$:
\begin{equation}
    \mathcal{PR}(V_{S(\mathcal{N},pi)},\pi) = \frac{V_h}{V_{S(\mathcal{N},\pi)}}
\end{equation}

\textbf{Remark 3.} \textit{Definition 7 allows us to quantify the performance of a certain control policy $\pi$ in faithfully carrying out safety specification $\mathcal{N}$. Imperfections of any control policy are expected, but it is the responsibility of safety verification to quantify and curb such imperfections.}

Practically, finding out about the specification penetration rate of a candidate controller for AVs in a specified ODD can be a challenge. First, deductive formal methods such as theorem proving have been limited to relatively simple discrete dynamic systems, and extending theorem proving to systems with continuous dynamics requires further established framework (a promising example \cite{immler2015verified}) and considerably higher computation resource. Second, traditional algorithmic formal methods such as model checking\cite{clarke1994model} also do not apply to systems with continuous dynamics, and running exhaustive tests of all possible actions would also create computation resource problem when fine discretization of continuous action space is needed. Recent progress in combining STL with reachability analysis\cite{chen2018signal} however, can be a possible solution to evaluate such penetration rate by calculating verification arithmetic on the candidate controller over the scenario space of the ODD, and predict the volume portion where the candidate controller will fail the STL specification. The specification penetration rate calculation problem is still an open one, and future work may combine different ways of evidence gathering, both deductive and algorithmic, to offer different alternatives.

\textbf{Definition 8. (Scenario safe coverage)} The scenario safe coverage $\mathcal{C}_\pi(\mathcal{O})$ of a policy $\pi$ to a scenario space in specified ODD $\mathcal{O}$ is defined as the ratio between safety verified volume and the full scenario space volume with the candidate policy $\pi$, either using sample-based verification or formal verification.

For simplicity, in the case of only one safety specification $\mathcal{N}$, the corresponding scenario coverage expressed in sample-based and formal methods are, respectively:
\begin{equation}
    \mathcal{C}_\pi(\mathcal{O}) = \frac{V_v{'} \cap V_{S(\mathcal{O})}}{V_{S(\mathcal{O})}},\quad 
    \mathcal{C}_\pi(\mathcal{O}) = \frac{V_h}{V_{S(\mathcal{N},\pi)}}
\end{equation}

\textbf{Remark 4.} \textit{In terms of scenario safe coverage, both sample-based methods or formal methods can provide confidence based on mathematical rigor at the simulation layer. Definition 6 is providing a unified framework for both school of methods to have level playing ground for comparison.}

The route to complete scenario coverage is illustrated in Figure \ref{fig:evolution}. Both formal and sample-based methods are valid tools to assist discovering AV safety control weakness, and AV control developers can select either or both route to their convenience and preference.

In summary, quantifying scenario coverage $\mathcal{C}_\pi(\mathcal{O})$ of a candidate AV control policy $\pi$ in a specified ODD $\mathcal{O}$ (similar to solving the safety quantification problem in \cite{weng2021}) can be performed by either sample-based method or formal methods, or even a mixture of both.

\subsection{Formal and Sample-based Approach Towards Full Scenario Coverage}
An illustrative example of the process for verifying scenario coverage in both formal and sample-based methods is provided in Figure \ref{fig:evolution}. A difference between the two methods, is that formal methods haven the inherent capability of determining a part of the scenario space where safety is infeasible, through for example reachability analysis.

\begin{figure}
    \centering
    \includegraphics[width=0.48\textwidth]{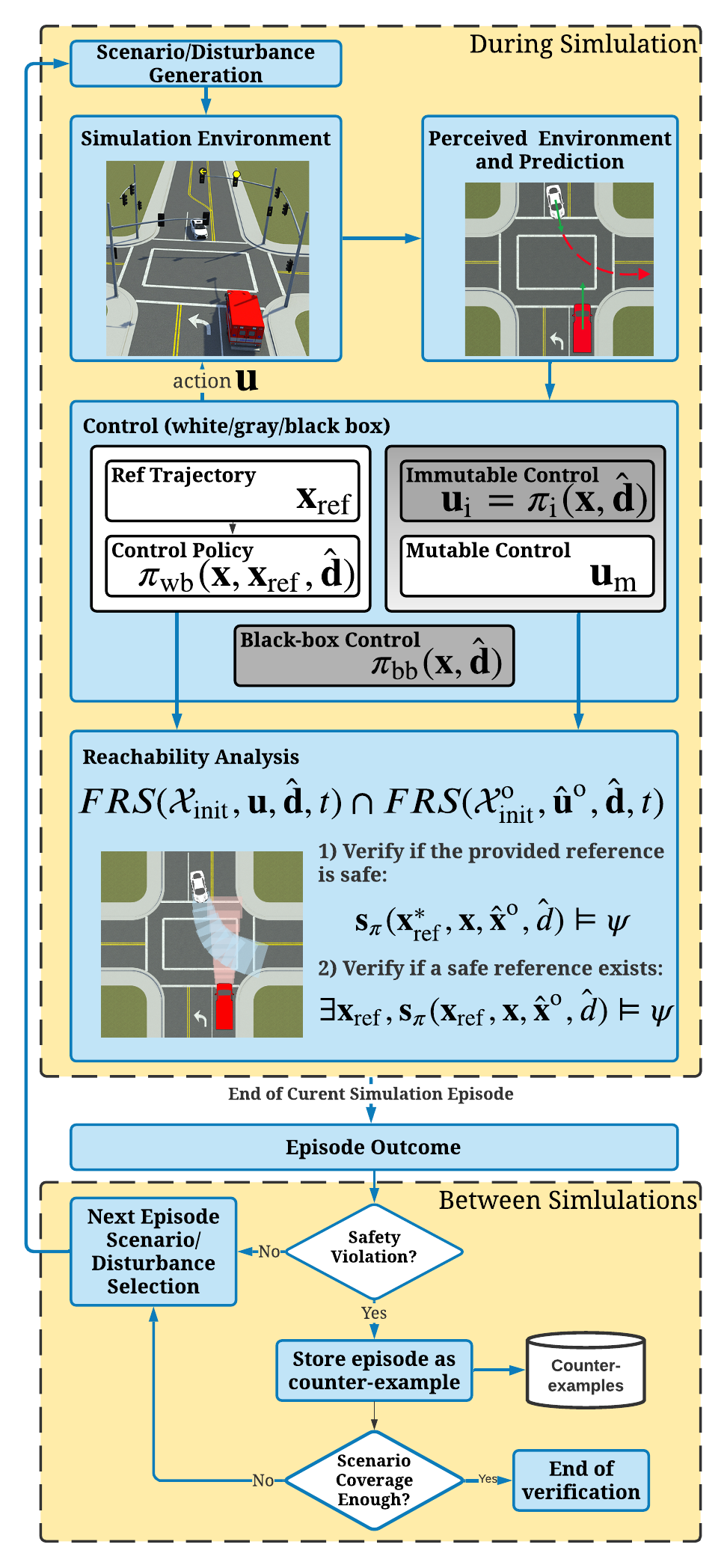}
    \caption{Procedure of different verification schemes: A priori verification can be performed for white-box and gray-box controls, shown in the top yellow box. Black-box control policy can only be verified through a posteriori verification, shown in the bottom yellow box. The difference is that a posteriori verification can only be performed between simulations using final safety outcome of each episode, while a priori verification can be performed as long as perception of environment is initiated in simulation.}
    \label{fig:control_verification}
\end{figure}

\section{State-of-the-art Formal Control Safety Verification}
Safety verification can be performed either during simulation (online) with a prediction module, or after simulation (offline) to simply check results (Figure \ref{fig:control_verification}). When the control policy of ego and participating traffic agents are known (\textit{white-box control}) or partially known (\textit{grey-bix control}), formal methods can leverage this information and narrow down the reachable sets to provide `tighter' prediction of vehicle motion, leading to valuable applications such as online safety monitoring and verification. Conversely, if the control policy is proprietary and completely unknown (\textit{black-box} control), the safety verification process will have to involve statistical learning of the black-box controller's behavior after each simulation, which in turn will guide the selection of next scenarios to be tested to better target counter-examples.

\subsection{Formal Safety Verification of Known Control Policy}
If the control policy is fully known (usually only limited to ego vehicle), then safety verification can be performed with much certainty, effectively removing the variable $u$ in the RA. This is usually done by computing reachable sets of controlled ego vehicle and comparing them with the occupancy set of obstacles in the interested time interval\cite{kochdumper2021verification}.

\subsection{Formal Safety Verification of Black-box Control Policy}

When the control policy is completely unknown, the verification needs to fall back to more general plans that do not require control policy information. Such plans treat the controller and the plant as an overall system, and investigate how the input (such as disturbance) to the overall system could lead to undesired output (such as safety violations). A survey on black-box verification\cite{corso2020survey} listed several such approaches as simulated annealing, evolutionary algorithms, Bayesian optimization, or extended ant colony optimization. 

\subsection{Formal Safety Verification of Grey-box Control Policy}
If the control policy is partially known, modified forms of reachability analysis can be performed.

In the setting of a semi-autonomous vehicle where the human driver's behavior is estimated at best, the safety verification has to be performed with partially known control policy (in this case the human driver policy). A particular lane keeping assist controller designed in this setting treats the human driver's steering policy as known and immutable, while treating the human driver's acceleration policy as unknown and mutable\cite{Falcone_2011}. Such treatment allows RA to predict the range of reachable states of vehicle within a proper time horizon, and the system can determine if the human driver needs assistance by comparing the current vehicle state with backward reachable safe sets.

When at least the qualitative objective of a control policy or function is known (e.g. activate intervention to save the vehicle from an avoidable collision due to driver mis-handling), then RA can be used to verify if such control policy or function has stayed faithful to its qualitative description. For example, RA can be used to tell if a collision avoidance system have missed or abused the opportunity to intervene\cite{nilsson2014verification}.

\subsection{Combine Formal Verification with Prediction}
To one extreme, if the worst-case assumption is made for other vehicles' behaviors in a dense traffic setting, which is that the participating vehicles are all wanting to run into the ego vehicle, then the ego vehicle will likely have a collision every a few seconds. This is easily a too conservative and useless assumption. To another extreme, if other traffic participants are assumed to be all perfectly rule obeying and acting like deterministic robots (e.g. driving at constant speed along their lane centers), then it is likely that the risk due to participant behavior uncertainty is underestimated. Prediction error can be removed if all participating vehicles communicate under vehicle-to-vehicle network\cite{kianfar2012reachability}, but this is often an ideal assumption. The problem of behavior prediction can be a topic of its own right, and often involves delicate distinctions of the information model and relates to game theory, where reachable sets become stochastic\cite{yoo2018predictive}. A list of behavior prediction algorithms used in the primary studies is provided in Table \ref{tab:behavior_prediction}. Presumptions of behaviors can lead to interesting guarantees regarding accident liability. As a particularly interesting case\cite{althoff_2014}, by modeling only traffic-obeying agents, it is claimed that legal safety can be guaranteed if ego vehicle trajectories are verified against such agents modelled as fully traffic-obeying, therefore no liability is attributed to ego vehicle. A similar assumption of ``responsibility-sensitive safety" (RSS) is proposed in Mobileye's whitebook for safety\cite{shalevshwartz2018formal}, assuming that other traffic participants should act according to some agreed ``common sense", and the AV should plan accordingly. Efforts of bridging the gap between reachability-based safety and RSS is seen in \cite{orzechowski2019towards}. Different subjective beliefs of ``aggressiveness" of other vehicles will affect the behavior prediction range, and subsequently affect the calculated safe reachable sets\cite{yoo2018predictive}. In \cite{li2021predictionbased} backward reachability is made less conservative by predicting human driving behavior and thus limiting action bounds of human driven vehicle.

\begin{table}[t]
    \centering
    \vspace{0.2cm}
    \caption{Behavior Prediction Methods}
    \begin{tabular}{p{.3\textwidth}p{.1\textwidth}}
        \hline\hline
        Method & References\\
        \hline
        \\
        Constant velocity vehicle & \cite{park2012model}\\
        Parameterized aggressiveness in lane-change behavior with Nash non-cooperative game model & \cite{yoo2018predictive}\\
        Perfect traffic-obeying agents (others) & \cite{althoff_2014}\\
        Longitudinal max acceleration constrained agents (others) & \cite{althoff_2009_collision_detection}\\
        Preview feedback controlled vehicle (ego) & \cite{Falcone_2011}\\
        Probabilistic motion pattern (defined as trajectory derivatives) & \cite{aoude2013GP_RRT_Reach}\\
        Classification between aggressive and non-aggressive driver using Martigale-based model & \cite{gao2019stochastic}\\
        Bayesian occupancy filter & \cite{Rummelhard_2015_CMCDOT,ledent2019formal}\\
        Responsibility-sensitive safety (Mobileye) & \cite{shalevshwartz2018formal}\\
        Multi-hypothesis filter & \cite{du2020online}\\
        
        \\
        \hline\hline\\
    \end{tabular}
    \label{tab:behavior_prediction}
\end{table}

\section{State-of-the-art Control Synthesis with Formal Safety Guarantee}
In the literature, formal safety guarantees have been developed for automated driving simulations and tests mainly by performing reachability analysis. The types of guarantees and associated reference in summarized in Table \ref{tab:forms_reachability}.

\begin{table*}[t]
    \centering
    \vspace{0.2cm}
    \caption{Forms of Reachability Analysis Deployment and Types of Guarantees Provided}
    \begin{tabular}{p{0.1\textwidth}p{0.3\textwidth}p{0.5\textwidth}}
        \hline\hline
        Reference & Form of Reachability Analysis & Type of Guarantee\\
        \hline
        \\
        \cite{sontges_2015,sontges_2017} &  maxFRS + minBRS & guarantees collision if no solution for ``anticipated reachable set" is found\\
        \cite{althoff_2011} & FRS of controlled dynamics & predicts unavoidable collision within control policy\\
        \cite{althoff_2007} & overapproximated maxFRS, maxFRT & all possible forward motions are included\\
        \cite{althoff_2009_collision_detection} & inaccuracy around trajectory-planned vehicle & checks probability of collision within the inaccuracy model bounds\\ 
        \cite{isoda_2007} & minBRT & guarantees collision free in horizon\\
        \cite{Falcone_2011} & maxBRS for partially-controlled \& uncontrolled vehicle & conservative collision-free guarantee for non-extreme driving\\
        \cite{vaskov2019not} & overapproximated maxFRI & guarantee ``not-at-fault" safety by checking potential collision in time intervals leading to end of time horizon\\
        \cite{nilsson2014safe} & maxBRT & guarantees safe interoperability: possibly safe human takeover from autonomous driving system\\
        \cite{kousik2017safe} & maxFRS of uncontrolled dynamics & guarantees collision free in horizon\\
        \cite{majumdar2017funnel} & funnel (a variant of FRT for controlled vehicle) & guarantees collision-free if a funnel can be found to stay clear of obstacles at all times\\
        \cite{ivanov2019verisig,ivanov2020verifying,ivanov2020case} & maxFRS implemented by Flow*\cite{chen2013flow} & guarantees safety of deep neural network (DNN) controlled close-loop system\\
        \cite{Orzechowski2018occulsion} & maxFRS of (possibly) occluded vehicle & guarantees collision-free with possibly occluded vehicle(s)\\
        \cite{tran2019decentralized} & maxFRS of each agent communicated through a decentralized network & real-time collision-free guarantee for a group of autonomous agents\\
        \cite{liebenwein2020compositional} & maxFRS + collision state pruning + maxBRS & guarantees collision-free and successful reach of target state if ``controller contract" (state constraint tubes) are honored\\
        \cite{du2020online} & maxFRT + pedestrian intent prediction & guarantees safety within a high probability bound of pedestrian motion\\

        \\
        \hline\hline\\
    \end{tabular}
    \label{tab:forms_reachability}
\end{table*}

\subsection{Formal Control Mode Arbitration}

Formal methods can provide useful information of the criticality of current driving situation. This information can be used to develop criteria for whether a different control scheme is needed to bring ego vehicle to safety from a safety-critical scenario.

One such arbitration criteria is whether the vehicle is in an \textit{inevitable collision state} (ICS)\cite{fraichard2004inevitable}, which some argue is a more robust measure of collision proximity than traditional threat metrics such as time-to-collision (TTC) in estimating collision frequency\cite{aasljung2016comparing}, a vital quantity in determining automotive safety integritity level (ASIL) in ISO26262\cite{ISO26262}. If the vehicle is indeed in such a state, an impact preparation should be executed (usually with extensive use of braking to slow down\cite{bouraine2011relaxing}, or by following \textit{not-at-fault} trajectories\cite{vaskov2019not}.) since collision is inevitable. This concept has been used to prototype ADAS functions such as determining when to trigger automatic emergency braking (AEB)\cite{savino2015triggering,yu2021autonomous} or producing collision avoidance maneuver\cite{martin_gomez2009}. To determine such ICS, reachability analysis is commonly used\cite{parthasarathi2007inevitable,Martinez-Gomez_2008,althoff2012lane,lawitzky2014determining}. When the prediction of moving obstacles is crucial in dynamic traffic scenarios, probabilistic framework should be adopted to create a probabilistic equivalent of ICS\cite{bautin_2010,althoff2010probabilistic}. For certain semi-autonomous or ADAS functions, the determination of whether system is in an inevitable collision state can help determine whether ADAS intervention is missed or uneccesary\cite{nilsson2014verification}.

Another arbitration criteria is whether target states are achievable under the current motion mode of the system. As an example, motion modes of a vehicle can be categorized using the notion of \textit{trim trajectories} characterized by the constant input required to maintain the motion\cite{frazzoli2000robust,frazzoli2002real}. A trajectory library of a certain system can be built for the controller to choose from\cite{majumdar2013robust,majumdar2017funnel}. Under such a context, \textit{continuous feasibility} and \textit{control liveliness} are key concerns ensuring the accomplishment of higher level tasks, such as overtaking another slow vehicle, merging into a busy fast lane, or avoiding an obstacle at high speeds\cite{barry2018high}. When selecting mode from a large library of complicated dynamic modes become computationally demanding, learning based approach such as reinforcement learning can be leveraged to efficiently learn the proper mode arbitration decision according to the environment\cite{williams2021trajectory,goddard2021utilizing}.

\subsection{Formal Methods for Direct Control Integration}

Reachability can also be directly integrated in the control synthesis ``at all times", providing continuous guidance/preference to the incumbent controller to avoid catastrophic events caused by infeasible control. 

\begin{figure}
    \centering
    \includegraphics[width=0.48\textwidth]{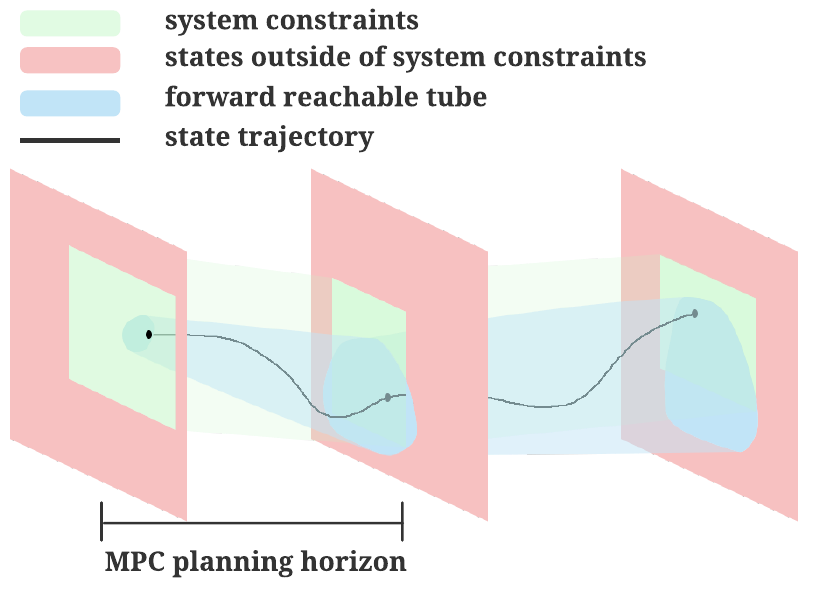}
    \caption{Robust MPC can take advantage of reachability to calculate a feasible set of states at the planning horizon (i.e. terminal state constraints), thus ensuring that the optimization process will not end up solving an infeasible constrained problem.}
    \label{fig:robust_mpc}
\end{figure}

One of such direct control integration methods is model predictive control (MPC) where formal methods especially reachability analysis can be used to develop robust MPC algorithms (Figure \ref{fig:robust_mpc}). Reachability analysis is used to find the feasible set of states for each MPC planning horizon so that resulting MPC algorithm will not try to find optimal solution in non-feasible region\cite{bemporad1999robust,bravo2006robust,dixit2018trajectory,rosolia2019sample}, leading to destabilization or control failure (Figure \ref{fig:robust_mpc}). This is also called recursive feasibility\cite{kvasnica2015reachability} or persistent feasibility\cite{kousik2020bridging}. When the reachable set constraints are probabilistic, a stochastic equivalent of robust MPC can be developed\cite{hewing2018stochastic,hewing2019scenario,gruber2019scalable}. Additional benefits of applying RA to MPC besides robustness and feasibility guarantee include a potential way of establishing system input-to-state stability\cite{limon2009input,raimondo2011robust}, separation of robustness and performance as in learning-based MPC\cite{aswani2013provably}, and reducing complexity in explicit MPC by removing regions that will never be reached\cite{kvasnica2019complexity}.

\begin{figure}
    \centering
    \includegraphics[width=0.48\textwidth]{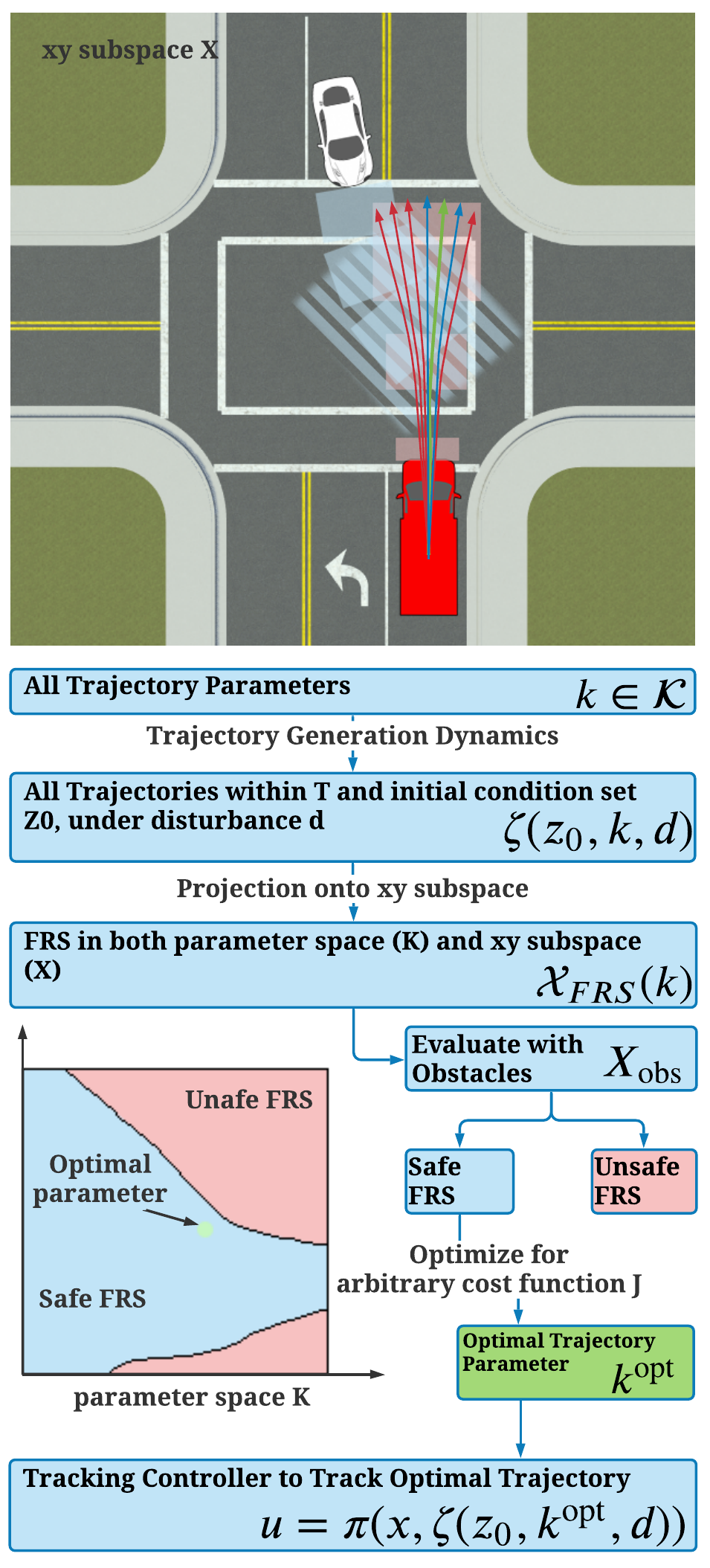}
    \caption{Reahability-based safe trajectory design process: first possible trajectories are parameterized (arrows in figure); then the trajectory parameters are combined with initial states to produce FRS (red shaded area) and evaluated with obstacle occupancy (blue shaded area for obstacle geometric center, blue striped area for occupancy at a certain time); the safe candidates are filtered and an optimal set of trajectory parameter is selected by a cost function; finally a lower level tracking controller tracks the optimal trajectory.}
    \label{fig:RTD}
\end{figure}

Another way to integrate formal safety into control design is through safe reference trajectory generation (Figure \ref{fig:RTD}). Reachability can be used in the trajectory planning phase to generate optimally safe and physically-feasible trajectories for lower level controller to follow\cite{kousik2017safe,vaskov2019guaranteed,kousik2020bridging}.

\section{Open Questions and Resources}

\subsection{Open Questions}
\subsubsection{Fidelity Speed Trade-off}
In performing reachability analysis, the trade-off between dynamics fidelity and computation tractability remains a major limit. In many occasions, full reachable set (reachable states with any possible control inputs) is neither necessary or economic. With a given control policy, the policy-bounded reachability can usually be computed in faster manner as a subset of the full reachable set. Moreover, the actual software/hardware implementation could further limit the configuration space, and by considering such limits into the RA, it is shown\cite{hildebrandt2020blending} that up to 91\% of computation load can be reduced.

In situations where actuation is less frequently demanded such as cruising on highway, computation can be further reduced by intermittently performing reachability analysis and skipping computation, while guaranteeing safety via control invariant set properties\cite{huang2020opportunistic}.

\subsubsection{Data-Driven Approach to Reachability}
An increasing number of studies are using data-driven approaches to address dynamical system verification problems such as estimation of dynamical system region of attraction\cite{kozarev2016case} or reachable set \cite{fenyes2018data,fenyes2020lpv,Driggs2018empiricalSet,alanwar2021data}. Dynamical systems can be approximated by deep neural networks(DNN), and to some extent the close-loop DNN-controlled system can be verified for safety by transforming DNN into equivalent hybrid system, and performing reachability analysis on the obtained hybrid system\cite{ivanov2019verisig,ivanov2020verifying,ivanov2020case}. As data-trained controllers are increasingly being tested and applied in autonomous systems such as aircraft collision avoidance system such as ACAS X\cite{kochenderfer2012next,julian2019deep}, the safety verification of neural network based controllers become research topic of interest. Feedforward neural network(FNN)\footnote{Feedforward neural network: neural network whose connections between nodes do not form a cycle, as opposed to recurrent neural networks, where connections form closed directed graph.} based closed-loop controllers using ReLU activation functions can be verified at their output layers using exact or overapproximated reachable sets to comply to simple sets of linear state constraints\cite{xiang2017reachable,xiang2018piecewiselinear,tran2019parallelizable,tran2019starset,yang2019efficient}. While the collision modeling in aircraft systems involve higher dimensions and susceptible to aerodynamic uncertainties, the scenarios are actually simpler and more straight forward. It is yet to see how such neural network based collision avoidance mechanism perform on automated ground vehicles, while some novel feedforward neural network training and verification using RA has been proposed for robotic systems\cite{chung2021constrained}. And if the collision avoidance modeling needs to be upgraded from its airborne counterpart, the methods to verify safety in \cite{xiang2017reachable,xiang2018piecewiselinear,tran2019parallelizable,tran2019starset,yang2019efficient} also need to be upgraded accordingly.

While vehicle behavior is largely governed by well studied vehicle dynamics, pedestrian behavior is much more unpredictable, and rely more on data-based estimation\cite{michael2018data,bansal2020hamilton,fisac2018probabilistically}, although some constraints such as velocity, acceleration or jerk can be assumed to curb the reachability\cite{hartmann2019pedestrians}.

\subsubsection{Beyond Binary Safety Verification}
Instead of using reachability for binary safety verification (e.g. safe or unsafe), it can also assist decision making in game theoretic traffic setting to increase ego vehicle payoffs\cite{meng2016dynamic}. Although collision-avoidance is a hard constraint that shall be met at all times to ensure safety, in a more general case, when different objectives of driving can be intermittently be compromised, the study of formal verification of weakly-hard systems\footnote{Weakly-hard system: systems that allow certain degree of constraint or deadline misses. The degree of allowed misses can be characterized by the ratio of missed versus all constraints/deadlines.} can become useful\cite{huang2019weaklyhard}. In fact, considering that human perception is flawed or even delayed up to seconds\cite{manassi2018serial}, weakly-hard systems is not such a bad compromise.

\subsubsection{Building New Logic For Safety Description}\label{sec:new_logic}
Although the majority of primary studies use linear temporal logic implicitly or explicitly for arguing safety guarantees, there exist developments of new forms of logic, such as \textit{Differential Dynamic Logic}\footnote{Differential dynamic logic(dL): a logic for specifying and verifying hybrid systems, verifiable with dL verification calculus, explained at \texttt{https://symbolaris.com/logic/dL.html}.}\cite{kolvcak2020relational}(based on concept in \cite{platzer2018logical}), allows statements and related proofs such as ``an earlier engagement of the emergency brake yields a smaller collision speed". It would be therefore interesting to see developments of new logic constructs for proving more complicated safety-related statements, as they could potentially help resolve liability issues generated by AVs.

\subsubsection{Ethical Verification with Reachability}\label{sec:ethics}
Ethical verification\cite{dennis2016formal} is an emerging field in automated vehicle decision making and planning. Some edge cases exist where zero-casualty is infeasible, and decision still have to be made based on ethical calculations\cite{dennis2013ethical}. Deontic logic\cite{hilpinen2012deontic}, which argues about the logic of duty and responsibility, is experimented to fit into traffic rule and accountability calculations\cite{rizaldi2015formalising,shea2021algorithmic,dennis2016formal}.

\subsection{Resources}

\subsubsection{Numerical Methods for Reachability Analysis}

Reachable sets originate from the evolution of set of states in system dynamics. Therefore there are two major system model based methods to approximate reachable sets numerically, i.e., set shape approximation\cite{bajcsy2019efficient,herbert2019warmstart} and system dynamics approximation\cite{althoff_2007,majumdar2017funnel,ivanov2019verisig}. Apart from model based methods, data sampling is another alternative to approximate reachable sets\cite{yel2019fast} or their control invariant subsets\cite{fenyes2018data,fenyes2020lpv}. Set contour approximation\cite{sontges_2015,sontges_2017,althoff_2016,Falcone_2011,xiang2017reachable,xiang2018piecewiselinear,tran2019parallelizable,yang2019efficient,girard_2005_zonotope,tran2019starset,herbert2019warmstart,bajcsy2019efficient}, dynamics approximation\cite{Asarin_2003_Girard,kousik2017safe,althoff_2010,hansen2011reachability,chen2013flow} and sample-based approximation\cite{Driggs2018empiricalSet,fenyes2018data,fenyes2020lpv,bak2021reachability,liebenwein2018sampling} are all viable numerical options. A summary of methods used in primary studies are listed in Table. \ref{tab:numerical_methods}. Academic competitions for reachability analysis exist, such as ARCH COMP\cite{ARCH20}, which is a competition of scientific software in the context of algorithmic verification of continuous and hybrid systems.

\begin{table*}
    \centering
    \vspace{0.2cm}
    \caption{Numerical Method for Reachable Set/Tube/Interval Calculation}
    \begin{tabular}{p{.2\textwidth}p{.25\textwidth}p{.25\textwidth}p{.2\textwidth}}
        \hline\hline
        Method & Pros & Cons & References\\
        \hline
        \\
        Rectangle approximation &  fast & accuracy suffers & \cite{sontges_2015,sontges_2017} \\
        Polytope approximation & relatively fast, convex/nonconvex & usually only convex shapes have been used to exploit fast calculations & \cite{althoff_2016,Falcone_2011,xiang2017reachable,xiang2018piecewiselinear,tran2019parallelizable,yang2019efficient}\\
        Zonotope approximation & fast\cite{girard_2005_zonotope} & conservative\cite{althoff_2008} and convex & \cite{althoff_2007,althoff_2008}\\
        Ellipsoid approximation & quadratic in nature, relatively fast, fits quadratic Lyapunov functions for calculation & shape may overly approximate vehicle occupancy & \cite{flus2020online}\\
        Star set approximation & Shape more flexible than zonotope or polytope, non-convex, less over-approximation & less intuitive  & \cite{tran2019starset}\\
        Piece-wise linear system equation approximation & good accuracy for nonlinear system approximation, proved convergence, preserves the region of limit cycles & convergence speed can be improved & \cite{Asarin_2003_Girard}\\
        Semi-definite programming over polynomial representations of continuous functions & has efficiency improvement potential & slow in computation speed, currently not real-time implementable & \cite{kousik2017safe}\\
        Differential inclusion of nonlinear dynamic model & lower computation complexity & accuracy may suffer from dynamic over-approximation, but partition techniques\cite{hansen2011reachability} can partially make up this shortcoming & \cite{althoff_2010,hansen2011reachability}\\
        Reachable set estimation using human-in-the-loop data sampling & fast since data sampling is offline, captures human driving pattern & data bias affects the validity of estimated set & \cite{Driggs2018empiricalSet}\\
        Controllable set estimation using simulation-generated data & high data coverage thanks to simulation & relies on heuristic classification criteria for controllability & \cite{fenyes2018data,fenyes2020lpv}\\
        Warm-start of HJI value function &  Guarantee over-approximation of true BRS, and reduces computation time compared to full initialization & the approximation to true BRS is not always exact, sometimes only conservative over-approximation can be achieved & \cite{herbert2019warmstart,bajcsy2019efficient}\\
        Local BRS update & significantly reduces computation time compared to full BRS update & limited to dealing with static obstacles in the environment & \cite{bajcsy2019efficient}\\
        Koopman operator linearization of data-driven model & Fast reachable set computation, system can be a black box model & relatively new approach & \cite{bak2021reachability}\\
        Taylor model(TM) based flowpipe tight over-approximation & Tight over-approximation to nonlinear ODEs, fast calculation & n/a & \cite{chen2013flow}\\
        Sample-based maxFRS under-approximation & The diversity of sampling is guaranteed by GreedyPack algorithm & GreedyPack is not the most efficient way to conduct diversified sampling & \cite{liebenwein2018sampling}\\
        Sample-based discrepancy function over-approximation & Works for black box dynamics model with white box transition graph & Subjects to learning bias & \cite{fan2017d,fenyes2020lpv}\\

        \\
        \hline\hline\\
    \end{tabular}
    \label{tab:numerical_methods}
\end{table*}

\subsubsection{Software Tools for Reachability Analysis}

For convenience, a set of software tools have been identified to perform reachability analysis. A list is provided in Table. \ref{tab:toolbox}. Apart from various calculation approaches, efforts in constructing benchmark for reachability tests\cite{chen2015benchmark} also help comparing effectiveness of different tools.

\begin{table*}[t]
    \centering
    \vspace{0.2cm}
    \caption{Software Tools for Reachability Analysis}
    \begin{tabular}{p{.15\textwidth}p{.4\textwidth}p{.38\textwidth}}
        \hline\hline
        Name & Website & Feature\\
        \hline
        \\
        Level-set Toolbox\cite{mitchell2005toolbox} &  \verb|https://www.cs.ubc.ca/~mitchell/ToolboxLS/| & Solves Hamilton-Jacobi-Issacs equation for reachability analysis using level-set approximation\\
        helperOC & \verb|https://github.com/HJReachability/helperOC| & An optimal control toolbox for Hamilton-Jacobi reachability analysis, a tutorial is available in the appendix of \cite{bansal2017hamilton}\\
        BEACLS & \verb|https://github.com/HJReachability/beacls| & A C++ implementation of the level-set methods for reachability analysis\\
        FaSTrack\cite{herbert2017FaSTrack} & \verb|https://github.com/HJReachability/fastrack| & Fast planning methods with slower, reachability-based safety guarantees for online trajectory planning in ROS framework\\
        KeYmaera X\cite{KeYmaeraX2017} & \verb|https://keymaerax.org/| & A theorem prover for differential dynamic logic\\ \\
        MPT\cite{mpt} & \verb|http://people.ee.ethz.ch/~mpt/2/| & Provides reachable set and invariance set calculation for linear, nonlinear and hybrid systems\\
        dReach\cite{kong2015dreach} & \verb|https://github.com/dreal/probreach| & A bounded reachability analysis tool for hybrid systems\\ \\
        SpaceEx\cite{frehse2011spaceex} & \verb|http://spaceex.imag.fr/| & A platform for the implementation of algorithms related to reachability and safety verification\\
        CommonRoad\cite{althoff2017commonroad} & \verb|https://commonroad.in.tum.de/| & a collection of composable benchmarks for motion planning on roads, which provides researchers with a means of evaluating and comparing their motion planners\\
        SPOT\cite{koschi2017spot} & \verb|http://koschi.gitlab.io/spot/| & A tool to predict the future occupancy of other traffic participants using reachable sets\\
        CORA & \verb|https://github.com/TUMcps/CORA| & A collection of MATLAB classes for the formal verification of cyber-physical systems using reachability analysis\\
        Flow*\cite{chen2013flow} & \verb|https://flowstar.org/| & A efficient tool to calculate reachability for polynomial-based hybrid systems\\
        ARIANDE & \verb|https://www.ariadne-cps.org/| & C++/Python library for formal verification of cyber-physical systems, using reachability analysis and rigorous numerics on nonlinear hybrid automata\\
        HyCreate\cite{bak2013hycreate} & \verb|http://stanleybak.com/projects/| \verb|hycreate/hycreate.html| & A tool for overapproximating reachability of hybrid automata using union of boxes to overapproximate reachable sets\\
        Hylaa\cite{bak2017hylaa} & \verb|http://stanleybak.com/hylaa/| & A verification tool for system models with linear ODEs, time-varying inputs, and possibly hybrid dynamics\\
        NNV & \verb|https://github.com/verivital/nnv/| & Implements reachability methods for analyzing neural networks, particularly with a focus on closed-loop controllers in autonomous cyber-physical systems (CPS)\\
        JuliaReach\cite{bogomolov_2019_JuliaReach} & \verb|https://juliareach.github.io/| & Reachability computations for dynamical systems in Julia\\ \\
        C2E2\cite{fan2016locally} & \verb|https://publish.illinois.edu/c2e2-tool/| & C2E2 can automatically check bounded time invariant properties of nonlinear hybrid automata\\
        DryVR\cite{fan2017d} & \verb|https://dryvrtool.readthedocs.io/en/latest/| &  Framework for probabilistic algorithm for learning sensitivity from simulation data, and bounded reachability analysis that uses the learned sensitivity\\
        AROC\cite{kochdumper2021aroc} & \verb|https://aroc.in.tum.de| & A toolbox to automatically synthesizes verified controllers for solving reach-avoid problems using reachability analysis\\

        \\
        \hline\hline\\
    \end{tabular}
    \label{tab:toolbox}
\end{table*}

\section{Conclusion}

In this paper we presented a holistic overview of the status quo in automated vehicle safety verification, validation and certification. In particular, the status and challenges of highly automated vehicle certification are pointed out, which should draw both academic and industrial attention as those challenges have a high reward as they are associated with major pain points in automated vehicle deployment. We identified a key missing link in quantifiable safety verification, which is a unified scenario coverage framework, and proposed such a definition to be expanded on by researchers of interest.

Formal safety verification is systematically over-viewed in this paper, including a state-of-art methodology review of its use in motion control safety verification and control synthesis with safety guarantee. Although reachability analysis, as a form of formal methods, have been used in multiple system control applications to guarantee robustness or constraint satisfaction, the lack of a unified and convenient engineering routine for safety specification penetration prevents the wider application of reachability analysis to more safety-critical functions in industry. With emerging papers starting to bridge the formal temporal logic expressions with dynamical system control synthesis, we suggested potential research opportunities, and listed relevant tools.



\section*{Acknowledgment}
The authors would like to thank Prof. Mo Chen at Simon Fraser University for providing high level feedback and suggestions.

\ifCLASSOPTIONcaptionsoff
  \newpage
\fi


\begin{IEEEbiography}[{\includegraphics[width=1in,height=1.25in,clip,keepaspectratio]{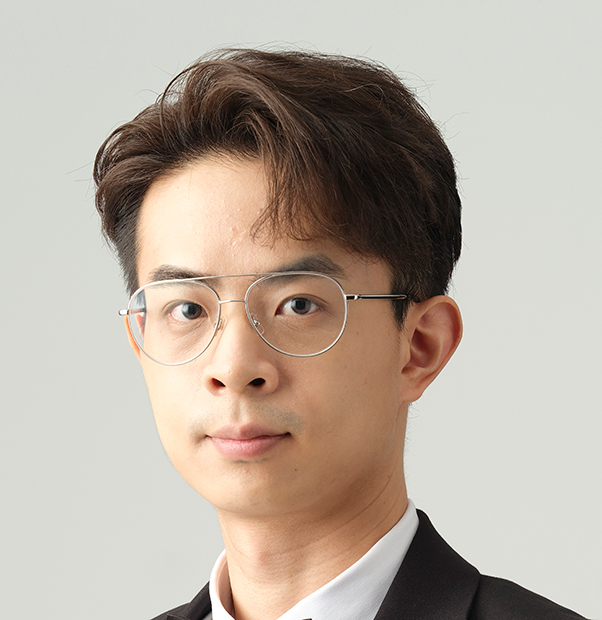}}]{Tong Zhao} 
(SM'17, M'21) received B.S. in Physics from Nanjing University, Nanjing, China and the University of Waterloo, Waterloo, Canada in 2012, and M.S in Electrical Engineering from Wuhan University, Wuhan, China in 2016. He is currently pursuing the Ph.D degree in Mechanical Engineering from the Ohio State University, Columbus OH.

In 2019 he worked as a control algorithm intern at tuSimple, Tucson, Arizona. In 2020 and 2021 he worked as software engineer intern at the Mathworks. His current research focus is on limit handling vehicle control algorithm and automated vehicle control safety guarantee through formal methods.
\end{IEEEbiography}

\begin{IEEEbiography}[{\includegraphics[width=1in,height=1.25in,clip,keepaspectratio]{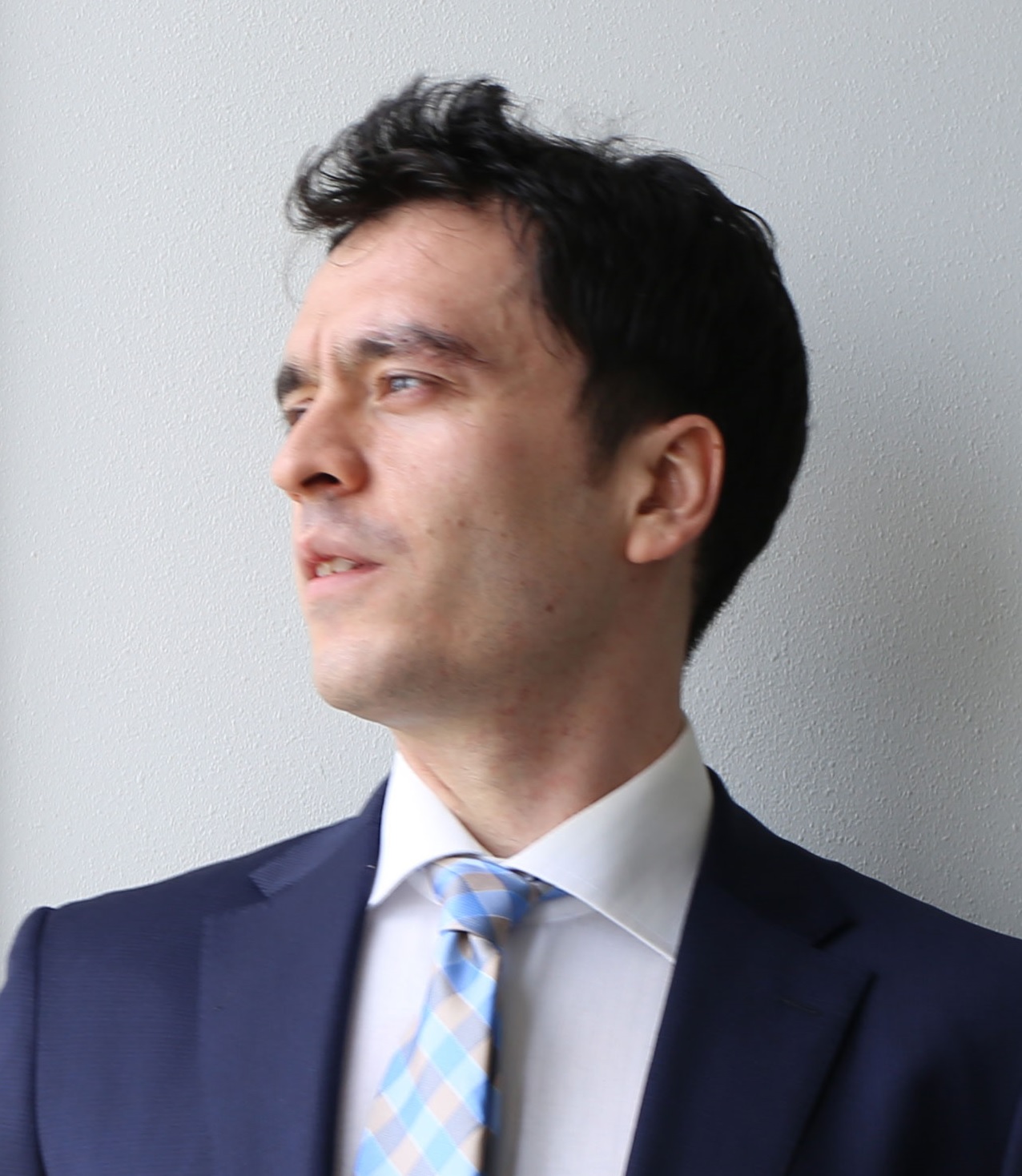}}]{Ekim Yurtsever} (Member, IEEE)
received his B.S. and M.S. degrees from Istanbul Technical University in 2012 and 2014 respectively. He received his Ph.D. in Information Science in 2019 from Nagoya University, Japan. Since 2019, he has been with the Department of Electrical and Computer Engineering, The Ohio State University as a research associate. 
	
His research focuses on artificial intelligence, machine learning, computer vision, reinforcement learning, intelligent transportation systems, and automated driving systems.

\end{IEEEbiography}

\begin{IEEEbiography}[{\includegraphics[width=1in,height=1.25in,clip,keepaspectratio]{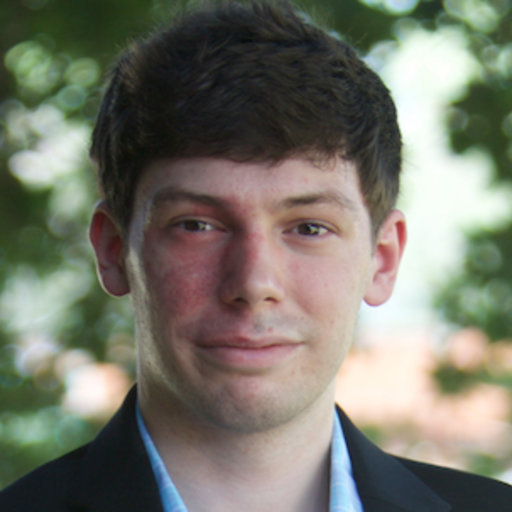}}]{Joel Paulson} received the B.S. degree from University of Texas Austin, Austin TX, in 2011, M.S.CEP and the Ph.D. degree in Chemical Engineering from Massachusetts Institute of Technology, Boston MA in 2013 and 2016, respectively.

From 2016 to 2019, he was a postdoc scholar at the Department of Chemical and Biomolecular Engineering, University of California, Berkeley CA. Since 2019, he has been an assistant professor in the Department of Chemical and Biomolecular Engineering at the Ohio State University. He focuses on the development of optimization, machine learning, and multi-scale simulation methods to improve the quality, efficiency, and sustainability of engineered products and processes. Two particular areas of focus in his group are: (i) construction of predictive models using a combination of physics-based equations and data and (ii) the formulation of real-time decision-support tasks under uncertainty as structured optimization problems that are solved efficiently using state-of-the-art algorithms. The developed strategies have been successfully applied to a broad range of systems including continuous pharmaceutical manufacturing, colloidal self-assembly, and non-equilibrium plasma jets.
\end{IEEEbiography}

\begin{IEEEbiography}[{\includegraphics[width=1in,height=1.25in,clip,keepaspectratio]{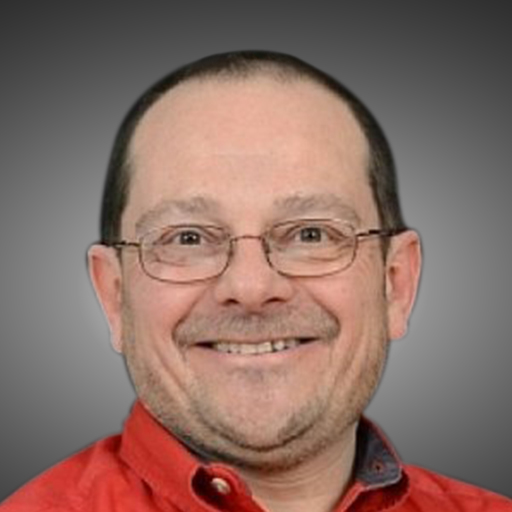}}]{Giorgio Rizzoni}
received B.S. (ECE) in 1980, M.S. (ECE) in 1982, Ph.D. (ECE) in 1986, all from the University of Michigan. 

Since 1999 he has been the director of the Ohio State University Center for Automotive Research (CAR), an interdisciplinary university research center in the OSU College of Engineering.  His research activities are related to modeling, control and diagnosis of advanced propulsion systems, vehicle fault diagnosis and prognosis, electrified powertrains and energy storage systems, vehicle safety and intelligence, and sustainable mobility. He has contributed to the development of graduate curricula in these areas, and has served as the director of three U.S. Department of Energy Graduate Automotive Technology Education Centers of Excellence: Hybrid Drivetrains and Control Systems (1998-2004), Advanced Propulsion Systems (2005-2011, and Energy Efficient Vehicles for Sustainable Mobility (2011-2016). Between 2011 and 2016 he served as the OSU Site Director for the U.S. Department of Energy China-USA Clean Energy Research Center - Clean Vehicles. He is currently leading an ARPA-E project in the NEXTCAR program.  During his career at Ohio State, Prof. Rizzoni has directed externally sponsored research projects funded by major government agencies and by the automotive industry in approximately equal proportion.  Prof. Rizzoni is a Fellow of SAE (2005), a Fellow of IEEE (2004), a recipient of the 1991 National Science Foundation Presidential Young Investigator Award, and of many other technical and teaching awards.
\end{IEEEbiography}


\end{document}